\documentclass{pasj00}
\usepackage[dvips]{color}
\draft

\begin{document}
\SetRunningHead{T. Nishimichi et al.}
{Characteristic Scales of Baryon Acoustic Oscillations from Perturbation 
Theory}
\Received{2007/05/02}
\Accepted{2007/08/??}

\title{Characteristic Scales of Baryon Acoustic Oscillations from 
Perturbation Theory: Non-linearity and Redshift-Space Distortion Effects}
\author{%
Takahiro \textsc{Nishimichi}\altaffilmark{1},
Hiroshi \textsc{Ohmuro}\altaffilmark{1},
Masashi \textsc{Nakamichi}\altaffilmark{2},
Atsushi \textsc{Taruya}\altaffilmark{1},\\
Kazuhiro \textsc{Yahata}\altaffilmark{1},
Akihito \textsc{Shirata}\altaffilmark{1,3},
Shun \textsc{Saito}\altaffilmark{1}, 
Hidenori \textsc{Nomura}\altaffilmark{2},\\
Kazuhiro \textsc{Yamamoto}\altaffilmark{2},
and,
Yasushi \textsc{Suto}\altaffilmark{1}}
\altaffiltext{1}
{Department of Physics, School of Science,
The University of Tokyo, Tokyo 113-0033, Japan}
\altaffiltext{2}
{Graduate School of Science, Hiroshima University, Higashi-Hiroshima, 
Hiroshima 735-8526, Japan}
\altaffiltext{3}
{Department of Physics, Tokyo Institute of Technology, 
Tokyo 152-8511, Japan}
\email{nishimichi@utap.phys.s.u-tokyo.ac.jp}
\maketitle
\begin{abstract}
An acoustic oscillation of the primeval photon-baryon fluid around the
decoupling time imprints a characteristic scale in the galaxy
distribution today, known as the baryon acoustic oscillation (BAO)
scale.  Several on-going and/or future galaxy surveys aim at detecting
and precisely determining the BAO scale so as to trace the expansion
history of the universe. We consider nonlinear and
redshift-space distortion effects on the shifts of the BAO scale in
$k$-space using perturbation theory. The resulting shifts are indeed
sensitive to different choices of the definition of the BAO scale, which
needs to be kept in mind in the data analysis. We present a toy model to
explain the physical behavior of the shifts.  We find that
the BAO scale defined as in Percival et al. (2007) indeed shows very
small shifts ($\lesssim 1\%$) relative to the prediction in
{\it linear theory} in real space. The shifts can be predicted accurately 
for scales where the perturbation theory is reliable.
\end{abstract}
\section{Introduction}
\label{sec:Intro}

The baryon acoustic oscillation (BAO) is an oscillation of photon-baryon
fluid imprinted in the matter spectrum as a characteristic signature.
Recently it was detected in the SDSS and 2dFGRS galaxy distribution
(e.g., \cite{Eisenstein2005,Hutsi2006,Tegmark2006,Padmanabhann2006,
Percival2007,Cole2005}), while its counterpart in the cosmic microwave
background (CMB) has already played an important role in precision
cosmology \citep{Spergel2007}.  Currently, using the BAO scale as a
standard ruler is regarded as one of the most promising tools to trace
the cosmic expansion history.  This characteristic scale basically
corresponds to the sound horizon at recombination (e.g.,
\cite{Eisenstein2005}, and see also Appendix \ref{app:phase}):
\begin{eqnarray}
r_s(z_{\rm rec}) &=& \int_{z_{\rm rec}}^{\infty}\,\frac{dz\,c_s(z)}{H(z)} 
\cr
&=& \frac{2}{3k_{\rm eq}}\sqrt{\frac{6}{R_{\rm eq}}}\ln
\frac{\sqrt{1+R_{\rm rec}}+\sqrt{R_{\rm rec}+R_{\rm eq}}}{1+\sqrt{R_{\rm
eq}}}
\cr
&\approx&
147(\Omega_mh^2/0.13)^{-0.25}(\Omega_bh^2/0.024)^{-0.08}{\rm Mpc},
\label{eq:soundhorizon}
\end{eqnarray}
where $c_s(z)$ is the sound speed at redshift $z$, and $z_{\rm rec}$ is
the redshift at recombination ($\simeq1089$). In the second equality,
$k_{\rm eq}$ is the horizon scale at the matter-radiation equality
epoch, $z_{\rm eq}$, $R_{\rm rec}=R(z_{\rm rec})$ and $R_{\rm eq}
=R(z_{\rm eq})$ are the ratio of the baryon to photon momentum densities
at $z_{\rm rec}$ and $z_{\rm eq}$. Finally the last equality is an
approximate fit where $\Omega_m$ and $\Omega_b$ are the density
parameters of matter and baryon, and $h$ is the current Hubble constant
in units of 100km$\,$s$^{-1}$Mpc$^{-1}$.

The BAO length scale itself can be computed accurately and is indeed
insensitive to the presence of dark energy that affects the expansion of
the universe at relatively low redshifts. Precisely for these reasons, 
it is a useful standard ruler of the universe. In particular, 
it is supposed to be a good tracer of $w_{\rm DE}\equiv p_{\rm DE}/
\rho_{\rm DE}$, where $p_{\rm DE}$ and $\rho_{\rm DE}$ are the pressure 
and the density of dark energy (e.g., \cite{Blake2003,Seo2003}). 
Also it can potentially be used to falsify the possibility of alternative 
law of gravity which might explain the acceleration of the cosmic expansion 
\citep{Shirata2005,Yamamoto2006}.

\begin{figure}[!ht]
   \centering \includegraphics[width=0.48\textwidth]{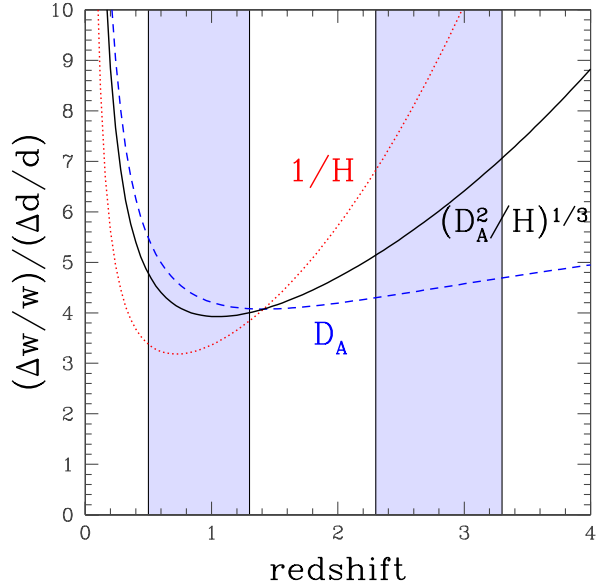}
\caption{The error propagation from measured scales, $d$, to the dark 
energy equation of state parameter, $w_{\rm DE}$, as a function of 
redshift. We choose $1/H(z)$ (dotted), and $D_A(z)$ (dashed) for $d$,
which correspond to the separations parallel and 
perpendicular to the line-of-sight direction. We also plot
the three dimensional average, $(D_A^2(z)/H(z))^{1/3}$ (solid) for $d$. 
The shaded regions indicate the targeted redshift 
ranges of a future galaxy survey, WFMOS.}
\label{fig:dwdr}
\end{figure}

Consider a sample of galaxies with measured redshifts, the observed BAO
scale provides estimates of the angular diameter distance $D_A(z)$ and
the inverse of the Hubble parameter $1/H(z)$, which correspond to the
scales perpendicular and parallel to the line-of-sight direction,
respectively. They in turn can be translated into the estimate of
$w_{\rm DE}$. Figure \ref{fig:dwdr} shows how the fractional errors of
three important scales, the angular diameter distance $D_{A}(z)$, the
inverse of Hubble parameter $1/H(z)$ and their average over three
dimensions $(D_A^2(z)/H(z))^{1/3}$ propagate to that of $w_{\rm
DE}$. The cosmological parameters assumed are the third year WMAP results 
(\cite{Spergel2007}, see section \ref{sec:Numerical} for details). 
The two shaded regions show the approximate targeted redshift ranges of a
future galaxy
redshift survey, WFMOS (Wide-field Fiber-fed Multi-Object Spectrograph).
Typically a ratio of $\Delta w/w$ and $\Delta d/d$ around $z=1$ ranges
from 3 to 5, while the value depends slightly on a specific choice
of cosmological parameters. Thus the $\sim3$\% determination of $w_{\rm
DE}$ requires the sub-percent accuracy/precision in determining the BAO
scale, which is challenging from observational, and even theoretical, 
points of view.

Until recently it was often assumed that gravitational non-linearity and
redshift-distortion effects do not shift the BAO scale, although 
they significantly affect the amplitude of the BAO. 
Some simulations (e.g., \cite{Meiksin1999,Seo2005,Springel2005,
Angulo2005,Jeong2006,Ma2006,Eisenstein2006,Angulo2007}) and analytical works 
(e.g., \cite{EisensteinSW2006,Guzik2007}; Smith, Scoccimarro \& Sheth 2007a; 
Smith, Scoccimarro \& Sheth 2007b) have been done 
to study these effects on the power spectrum or two-point correlation 
function at the BAO scale. 
For a few percent accuracy on $w_{\rm DE}$, more precise BAO predictions are 
required, which we consider in the present paper on the basis of the 
one-loop corrections from perturbation theory.

Very recently, several authors (Crocce \& Scoccimarro 2006a; Crocce \&
Scoccimarro 2006b; \cite{McDonald2007,Valageas2007,Padmanabhan2006};
Matarrese \& Pietroni 2007a; Matarrese \& Pietroni 2007b; \cite{Crocce2007}) 
attempt to take account of higher-order corrections. For the first time, we 
take account of the redshift-space distortion effects on the BAO scale 
applying the result of \citet{Scoccimarro2004}.

The outline of this paper is as follows; section \ref{sec:Formulation} 
briefly summarizes the perturbation theory of density and velocity 
fields. The details of the numerical calculation are shown in section
\ref{sec:Numerical}. The results are shown in section 
\ref{sec:Results} with discussion and interpretation in 
section \ref{sec:Interpretation}. Finally, section 
\ref{sec:Conclusion} is devoted to the conclusion of this paper.

\section{Non-linear power spectrum in real and redshift spaces}
\label{sec:Formulation}

Cosmological perturbation theory predicts 
the gravitational clustering of matter distribution 
by a systematic expansion of the cosmological 
density and velocity fields.  
It provides us an accurate prediction 
to the non-linear modification of matter power spectrum  
as long as the gravitational clustering is mildly non-linear 
(e.g., \cite{Juszkiewicz1981,Vishniac1983,Fry1984,Goroff1986,Suto1991,
Makino1992,Jain1994,Bernardeau1994,Matsubara1995,
Scoccimarro1998,Chodorowski2002}, and see \cite{Bernardeau2002} for a review). 
In this section, we briefly review the formalism to calculate the 
non-linear correction to the matter power spectrum. 
A model of redshift-space distortion is also presented, which will 
be used later to estimate the non-linearity of redshift-space power 
spectrum.

\subsection{Perturbation theory}
\label{subsec:PT}

To deal with the gravitational clustering of the matter distribution,  
we adopt the hydrodynamic description and treat the dark matter 
and baryons as a pressureless dust fluid. 
According to \citet{Scoccimarro2001}, 
the evolution equations for the cosmic fluid can be written 
in a compact form by introducing the two-component vector: 
\begin{eqnarray}
\Psi_i({\bf k};\eta) \equiv \left(\delta({\bf k};\eta),
-\frac{\theta({\bf k};\eta)}{{\cal H}(\eta)f(\eta)}\right),
\end{eqnarray}
where the subscript $i=1,\,2$ selects the density or velocity components, 
with $\delta({\bf k})$ and 
$\theta({\bf x};\eta)\equiv\nabla\cdot{\bf v}({\bf x};\eta)$ 
respectively being the Fourier transform of the density fluctuation 
and peculiar velocity divergence. 
The variable $\eta$ represents the time variable 
defined by $\eta\equiv\ln D_+$, where 
$D_+$ is the linear growth factor normalized to unity at present. 
The quantity ${\cal H}$ is the conformal 
expansion rate given by ${\cal H}\equiv d\ln a/d\tau$, where  
$a$ is the cosmic scale factor, and the quantity $f$ is the  
logarithmic derivative of the linear growth 
factor, $f(\eta)\equiv d\ln D_+/d\ln a$.

Assuming the irrotational fluid flow, the evolution equations 
can then be written as 
\begin{eqnarray}
\partial_\eta\Psi_i({\bf k};\eta)+\Omega_{ij}(\eta)
\Psi_j({\bf k};\eta)=\int d^3k_1d^3k_2\gamma_{ijk}
({\bf k},{\bf k}_1,{\bf k}_2)\Psi_j({\bf k}_1;\eta)\Psi_k({\bf k}_2;\eta),
\label{eq:EOMforPsi}
\end{eqnarray}
where the kernel $\gamma_{ijk}$ is the vertex matrix which represents 
the non-linear interaction between different Fourier modes: 
\begin{eqnarray}
\gamma_{ijk}({\bf k},{\bf k}_1,{\bf k}_2) &\equiv& \left\{
\begin{array}{ll}
\delta_D({\bf k}-{\bf k}_1-{\bf k}_2)\frac{\displaystyle
({\bf k}_2+{\bf k}_1)\cdot{\bf k}_2}{\displaystyle 2\,k_2^2}
&{\rm for}\quad(i,j,k)=(1,1,2),\\
\delta_D({\bf k}-{\bf k}_1-{\bf k}_2)\frac{\displaystyle
({\bf k}_1+{\bf k}_2)\cdot{\bf k}_1}{\displaystyle 2\,k_1^2}
&{\rm for}\quad(i,j,k)=(1,2,1),\\
\delta_D({\bf k}-{\bf k}_1-{\bf k}_2)\frac{\displaystyle
|{\bf k}_1+{\bf k}_2|^2
({\bf k}_1\cdot{\bf k}_2)}{\displaystyle 2\,k_1^2k_2^2}
&{\rm for}\quad(i,j,k)=(2,2,2),\\
0&{\rm otherwise},
\end{array}
\right.
\end{eqnarray}
with $\delta_D$ being Dirac's delta function. The matrix $\Omega_{ij}$ 
is given by 
\begin{eqnarray}
\Omega_{ij}(\eta) &\equiv& \left[
\begin{array}{cc}
0&-1\\
 -3/2&1/2
\end{array}
\right]. 
\end{eqnarray}
This expression is not exact in the non Einstein-de Sitter cases,  
but it still provides an accurate prescription.

The evolution equation given above is systematically 
solved by a perturbative expansion of 
two-component vector. Ignoring the decaying modes, one obtains 
\begin{eqnarray}
\Psi_i({\bf k};\eta) &=& \sum_{n=1}^{\infty}e^{n\eta}\psi_i^{(n)}({\bf k});
\label{eq:perturb_sol}
\\
\psi_i^{(n)} &=& \int \frac{d^3k_1\cdots d^3k_n}{(2\pi)^{3n-3}}
\delta_D({\bf k}-{\bf k}_1-\cdots
-{\bf k}_n){\cal F}_i^{(n)}({\bf k}_1,\cdots,{\bf k}_n)\,
\delta_1({\bf k}_1)\cdots\delta_1({\bf k}_n),
\label{eq:Kernel}
\end{eqnarray}
where the quantity $\delta_1$ stands for 
the initial density fluctuation, which we assume is a Gaussian random variable.
The Fourier kernel ${\cal F}_i^{(n)}({\bf k}_1,\cdots,{\bf k}_n)$ 
represents the mode coupling between different Fourier modes originating 
from the non-linear interactions. In Appendix \ref{app:Kernel}, 
the explicit expressions for ${\cal F}_i^{(n)}$ are presented 
up to the third-order of perturbative expansion.

We are interested in the non-linear evolution of power spectrum, which   
are evaluated as the ensemble average of the quantity $\Psi_i$: 
\begin{eqnarray}
\langle\Psi_i({\bf k};\eta)\Psi_j({\bf k'};\eta)\rangle = (2\pi)^3
\delta_D({\bf k}+{\bf k'})P_{ij}(k;\eta)\quad\quad(i,j=1,2).
\end{eqnarray}
Note that we obtain the three different power spectra: $P_{\delta\delta}$ 
from $(i,j)=(1,1)$, $P_{\delta\theta}$ from $(i,j)=(1,2)$ and $(2,1)$, 
and $P_{\theta\theta}$ from $(i,j)=(2,2)$. 
Substituting the solutions (\ref{eq:perturb_sol}) up to 
the third-order perturbations into the 
above, the next-to-leading order corrections are obtained, which can be 
summarized as   
\begin{eqnarray}
P_{ij}(k;\eta) &=& D_+^2P^{\rm L}(k)+D_+^4
\left[P_{ij}^{(13)}(k)+P_{ij}^{(22)}(k)\right]. 
\label{eq:1loop}
\end{eqnarray}
In the above expression, the first term in the right-hand-side is the 
linear power spectrum given by 
\begin{eqnarray}
\langle\delta_1({\bf k})\delta_1({\bf k'})\rangle
\equiv(2\pi)^3\delta_D({\bf k}+{\bf k'})P^{\rm L}(k). 
\end{eqnarray}
On the other hand, the terms in the bracket of equation 
(\ref{eq:1loop}) are the so-called one-loop corrections as a result of 
the non-linear mode-coupling:  
\begin{eqnarray}
P_{ij}^{(13)}(k) &\equiv& 3\int \frac{d^3q}{(2\pi)^3}
\left[{\cal F}_i^{(3)}({\bf k},{\bf q},-{\bf q})+
{\cal F}_j^{(3)}({\bf k},{\bf q},-{\bf q})\right]P^{\rm L}
(k)P^{\rm L}(q),
\label{eq:P13}\\
P_{ij}^{(22)}(k) &\equiv& 2\int \frac{d^3q}{(2\pi)^3}
\left[{\cal F}_i^{(2)}({\bf k}-{\bf q},{\bf q})
{\cal F}_j^{(2)}({\bf k}-{\bf q},{\bf q})\right]P^{\rm L}
(|{\bf k}-{\bf q}|)P^{\rm L}(q).
\label{eq:P22}
\end{eqnarray}
Note that the part of the integrals over the Fourier mode ${\bf q}$ are 
analytic, which we use. The resultant expressions include 
the two-dimensional integrals over $r$ and $x$ 
for the $P_{ij}^{(22)}(k)$ part, while 
the $P_{ij}^{(13)}(k)$ part has the one-dimensional integral over $r$.  
Their explicit functional forms are summarized in 
Appendix \ref{app:Kernel}, together with the definition of $r$ and $x$.

\subsection{A Model of redshift-space distortion}
\label{subsec:RD}

The observed distribution of galaxies constructed from redshift surveys 
is inevitably distorted due to the peculiar velocity of each galaxy. 
This effect, known as redshift-space distortion, is classified in 
two ways. On large scales, the bulk motion falling into a cluster 
apparently squashes the matter distribution, which enhances the 
clustering signal along the line-of-sight. 
On small scales, on the other hand, 
the virialized random motion of galaxies residing 
at a cluster suppresses the amplitude of the clustering signal. 
This is called the fingers-of-God (FOG) effect.

Based on the linear perturbation theory, 
\citet{Kaiser1987} proposed a formula for redshift-space power 
spectrum on large-scales: 
\begin{eqnarray}
P^{(\rm s)}(k,\mu;\eta) = \left[1+f(\eta)\mu^2\right]^2P^{(\rm r)}(k;\eta),
\label{eq:Kaiser}
\end{eqnarray}
where $\mu$ is the cosine of the angle between the line-of-sight direction 
and the Fourier mode $\mathbf{k}$. The spectra 
$P^{(\rm r)}(k;\eta)$ and $P^{(\rm s)}(k,\mu;\eta)$ respectively denote 
the matter power spectra in real and redshift spaces. 
Several authors proposed models for redshift-space power spectrum 
taking account of the small-scale random motion 
\citep{Peacock1994,Park1994,Cole1994,Ballinger1996,Magira2000}. 
In their models, the FOG effect 
is expressed by a damping factor, $D_{\rm FOG}(k,\mu)$, which they assumed  
Gaussian or Lorentzian. Their models are written as
\begin{eqnarray}
P^{(\rm s)}(k,\mu;\eta) = \left[1+f(\eta)\mu^2\right]^2P^{(\rm r)}
(k;\eta)D_{\rm FOG}(k,\mu).
\label{eq:RD1}
\end{eqnarray}
Note that the damping factor $D_{\rm FOG}(k,\mu)$ asymptotically approaches 
unity in the large-scale (small $k$) limit.

We should notice that the models given in equation (\ref{eq:RD1}) 
have been constructed to deal with a relatively small-scale clustering. 
It might not be accurate enough for our interest in the precision measurement 
on BAO scales. As for the models (\ref{eq:Kaiser}), it has been advocated 
that nonlinear random motion can not be negligible 
even in the large-scale limit and one could 
not recover the Kaiser formula (\ref{eq:Kaiser}) \citep{Scoccimarro2004}. 
We therefore look for alternative models 
relevant for the BAO scales.

\citet{Scoccimarro2004} recently proposed a 
physically plausible model of redshift-space distortion 
based on perturbation theory. 
He improved the Kaiser formula to take account of the non-linear 
evolution of density and the velocity fields, as well as the 
FOG effect. Using the three different power spectra defined in equation 
(\ref{eq:1loop}),  the explicit expression for the redshift-space spectrum 
becomes 
\begin{eqnarray}
P^{(\rm s)}(k,\mu;\eta) &=& \left[P_{\delta\delta}(k;\eta)+2f(\eta)\mu^2 
P_{\delta\theta}(k;\eta)+f(\eta)^2\mu^4P_{\theta\theta}(k;\eta)\right]
\nonumber\\
&&\times\exp\left[-f(\eta)^2\mu^2k^2\sigma_v^2(\eta)\right],
\label{eq:Scoccimarro}
\end{eqnarray}
with the quantity $\sigma_v^2(\eta)$ being the one-dimensional 
linear velocity dispersion: 
\begin{eqnarray}
\sigma_v^2(\eta) = \frac{1}{3}\int\frac{d^3k}{(2\pi)^3}
\frac{P_{\theta\theta}(k;\eta)}{k^2}.
\label{eq:sigmav2}
\end{eqnarray}
The model (\ref{eq:Scoccimarro}) properly accounts for the 
non-linear mode-couplings of density-density, density-velocity and 
velocity-velocity fields in the Kaiser formula. 
In this paper, we adopt the model in equation (\ref{eq:Scoccimarro}) 
to calculate the redshift-space power spectrum. Note 
that our current model of the FOG effect is still empirical and has to 
be tested with numerical simulations, which will be discussed in future 
work.

To compute $P^{(\rm s)}$, we use the one-loop results of power spectrum 
except for the linear velocity dispersion (\ref{eq:sigmav2}). The 
results are  
then presented by taking the angular average:  
\begin{eqnarray}
P^{(\rm s)}_0(k;\eta) &=& \frac{1}{2}\int_{-1}^{1}d\mu 
P^{(\rm s)}(k,\mu;\eta).
\end{eqnarray}

\section{Details of calculation} 
\label{sec:Numerical}

\subsection{Initial condition}
\label{subsec:Initial}

In what follows, we use the {\tt CAMB} code \citep{Lewis2000} to 
calculate the transfer function for the linear power spectrum 
$P^{\rm L}(k)$. We assume a flat $\Lambda$CDM model with adiabatic 
Gaussian fluctuation. We use the best-fit values of the cosmological 
parameters determined from the three year WMAP data \citep{Spergel2007}: 
$\Omega_mh^2=0.1277$, $\Omega_bh^2=0.02229$, $h=0.732$, $n_s=0.958$, 
$\tau=0.089$ and $\sigma_8=0.761$, where $n_s$ is the scalar 
spectral index (without running), 
$\tau$ is the optical depth and $\sigma_8$ is 
the rms of the density contrast smoothed with an 8$h^{-1}$Mpc top-hat 
window. Note that from equation (\ref{eq:soundhorizon}), we have 
$r_s=148$Mpc, which leads to the 
oscillatory behavior like $\sin(kr_s)$ in the transfer function 
of the matter fluctuation.

\subsection{Accuracy of the perturbation prediction}
\label{subsec:limit}

In general, the prediction from perturbation theory eventually breaks down 
when the non-linear correction dominates over the linear theory 
prediction. While we cannot rigorously define the validity range of 
the perturbation prediction, \citet{Jeong2006} recently showed 
that the one-loop correction to the power spectrum $P_{\delta\delta}(k)$ 
accurately describes the N-body simulations to better than 
1\% accuracy when 
\begin{eqnarray}
\Delta^2(k;z) \equiv \frac{k^3P_{\delta\delta}(k;z)}{2\pi^2} 
\lesssim
0.4. 
\label{eq:limit_PT}
\end{eqnarray}
Thus, for the limitation of the perturbation predictions, 
we define the maximum wavenumber $k_{\rm 1\%}$ through 
\begin{eqnarray}
\Delta^2(k_{\rm 1\%}(z);z) = 0.4. 
\label{eq:knl}
\end{eqnarray}
Note that the limitation for the power spectra $P_{\delta\theta}$ and 
$P_{\theta\theta}$ would not be described by equation (\ref{eq:limit_PT}) 
(see Crocce \& Scoccimarro 2006b). Nevertheless, for simplicity,  
we adopt the above criterion in both real and redshift spaces.

\begin{figure}[!ht]
   \centering \includegraphics[width=0.48\textwidth]{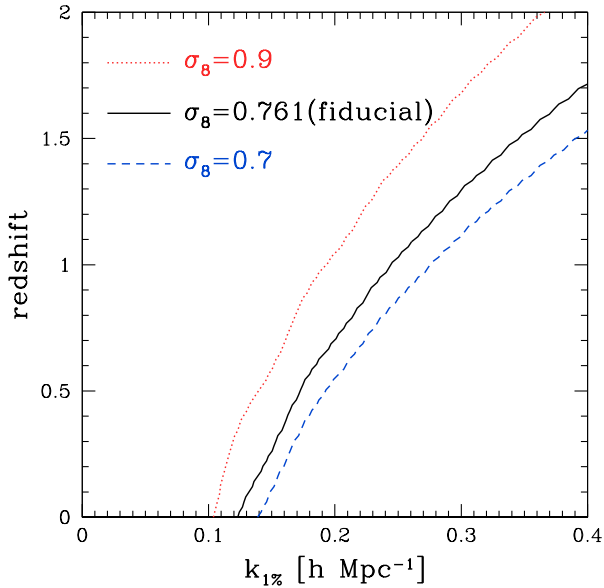}
\caption{The maximum wavenumber of the validity range for the perturbation 
theory, $k_{\rm 1\%}$, defined by equation (\ref{eq:knl}). The solid 
line represents the result for our fiducial model. We also plot the results 
for the cases with a slightly larger amplitude ($\sigma_8=0.9$; dotted), 
and a smaller amplitude ($\sigma_8=0.7$; dashed).}
\label{fig:knl}
\end{figure}

Figure \ref{fig:knl} shows the numerical values of $k_{\rm 1\%}$ 
as a function of redshift. Here, in addition to the result for 
our fiducial cosmological model (solid), we also plot the results for 
the cases with slightly different amplitudes: $\sigma_8=0.7$ (dashed) 
and $0.9$ (dotted), keeping the other remaining parameters fixed. 
The validity range of perturbation theory has a strong dependence on 
the redshift and the normalization. For the higher fluctuation 
amplitude with $\sigma_8=0.9$, the perturbation theory breaks down 
at a relatively smaller wave number.

\subsection{Characterizing the acoustic oscillation scales}
\label{subsec:scale}

We are especially concerned with the systematic influences of 
the non-linear clustering and the redshift-space distortion on 
the characteristic scale of BAO, affecting its use  as a cosmic 
standard ruler. To quantify the degree of these influences, 
we must first define the algorithm to characterize the 
oscillatory features from the power spectrum, by which  
the locations of peaks and the troughs are identified unambiguously.  
Then, the systematic influences can be investigated by measuring 
both the amplitude and the locations of peaks (and troughs). 
There are several approaches to 
separate the oscillatory pattern from the power-law behavior of 
power spectrum. In this paper, we study the following three methods:

\begin{description}

\item[(i)] Divide the matter power spectrum by a smooth linear spectrum: 
\begin{eqnarray}
f^{\rm (i)}_{\rm BAO}(k) = \frac{P(k)}{P_{\rm nw}(k)},
\label{eq:f1}
\end{eqnarray}
where $P_{\rm nw}(k)$ represents the ``no-wiggles'' approximation of the 
linear power spectrum given by \citet{Eisenstein1998}, which can be 
evaluated for our fiducial cosmological model. 
With this characterization, smooth power-law behavior 
in the power spectrum is effectively eliminated and the peaks and the 
troughs of acoustic oscillations are clearly identified. 
Note, however, that the above definition only corrects the 
power-law feature of the linear spectrum and non-linear corrections 
of power spectrum are left untouched in the numerator.  
Furthermore, there is uncertainty in the normalization of the 
amplitude.   

\item[(ii)] Take the logarithmic derivative:
\begin{eqnarray}
f^{\rm (ii)}_{\rm BAO}(k) = \frac{d\ln P(k)}{d\ln k}.
\label{eq:f2}
\end{eqnarray}
In contrast to the method (i), this method is free from 
the uncertainty in the normalization 
of amplitude. Also, it automatically separate the smooth power-law 
feature of $P(k)$ from the oscillatory behavior. 
In practice, however, it seems rather difficult in estimating  
the logarithmic derivative from the noisy binned data.

\item[(iii)] Divide the matter power spectrum by the smooth spectrum 
constructed from the matter power spectrum itself: 
\begin{eqnarray}
f^{\rm (iii)}_{\rm BAO}(k) = \frac{P(k)}{P_{\rm smooth}(k)}.
\label{eq:f3}
\end{eqnarray}
This method is almost identical to the one proposed by \citet{Percival2007}. 
Since we do not need a reference spectrum, it seems 
useful in characterizing the acoustic signature from 
the observed spectrum. 
Here, the smooth spectrum is constructed as follows. First we sample 
70 data points from our perturbation spectrum between $k=0.02h$Mpc$^{-1}$ 
and $0.3h$Mpc$^{-1}$ with equal separation and without weight. 
Then we select $8$ nodes separated by $\Delta k=0.05h$Mpc$^{-1}$ between 
$0.025h$Mpc$^{-1}$ and $0.375h$Mpc$^{-1}$, and an additional node at 
$k=0.001h$Mpc$^{-1}$. Connecting the sample points by fitting cubic B-spline 
functions at each node, we obtain the smooth power spectrum $P_{\rm
smooth}(k)$. 
\end{description}

Finally, evaluating the function $f_{\rm BAO}(k)$ in each method, 
we search for the local maxima and minima,  
which we respectively call {\it peaks} and 
{\it troughs} to characterize the sound horizon scales in BAOs: 
\begin{eqnarray}
\left.\frac{d\,f_{\rm BAO}(k)}{dk}\right|_{\rm peak\;or\;trough} = 0.  
\label{eq:def_peak_trough}
\end{eqnarray}

As a reference, positions of peaks and 
troughs are computed in each method for the linear power 
spectrum and the resultant numerical values are 
listed in Table \ref{tab:linear_position}.

%
%
\begin{table}[!t]
\caption{The linear theory predictions of peak and trough positions in 
units of $h$Mpc$^{-1}$ for our three methods. 
}
\begin{center}
\begin{tabular}{|l||c|c|c|c|c|c|c|c|c|}
\hline
  method & 1st peak & 2nd peak & 3rd peak & 4th peak & 1st trough &
  2nd trough & 3rd trough & 4th trough \\ 
\hline
(i) $P/P_{\rm nw}$ & 0.0669 & 0.1221 & 0.1792 & 0.2377 & 
0.0415 & 0.0944 & 0.1500 & 0.2075
\\ \hline
(ii) $d\ln P/d\ln k$ & 0.0533 & 0.1074 & 0.1638 & 0.2216 &
0.0823 & 0.1368 & 0.1933 & 0.2514
\\ \hline
(iii) $P/P_{\rm smooth}$ & 0.0663 & 0.1224 & 0.1788 & 0.2370 & 
0.0417 & 0.0940 & 0.1506 & 0.2076
\\ \hline
\end{tabular}
\end{center}
\label{tab:linear_position}
\end{table}

\section{Baryon Acoustic Oscillations from perturbation theory}
\label{sec:Results}

\begin{figure}[t]
   \centering \includegraphics[width=0.4\textwidth]{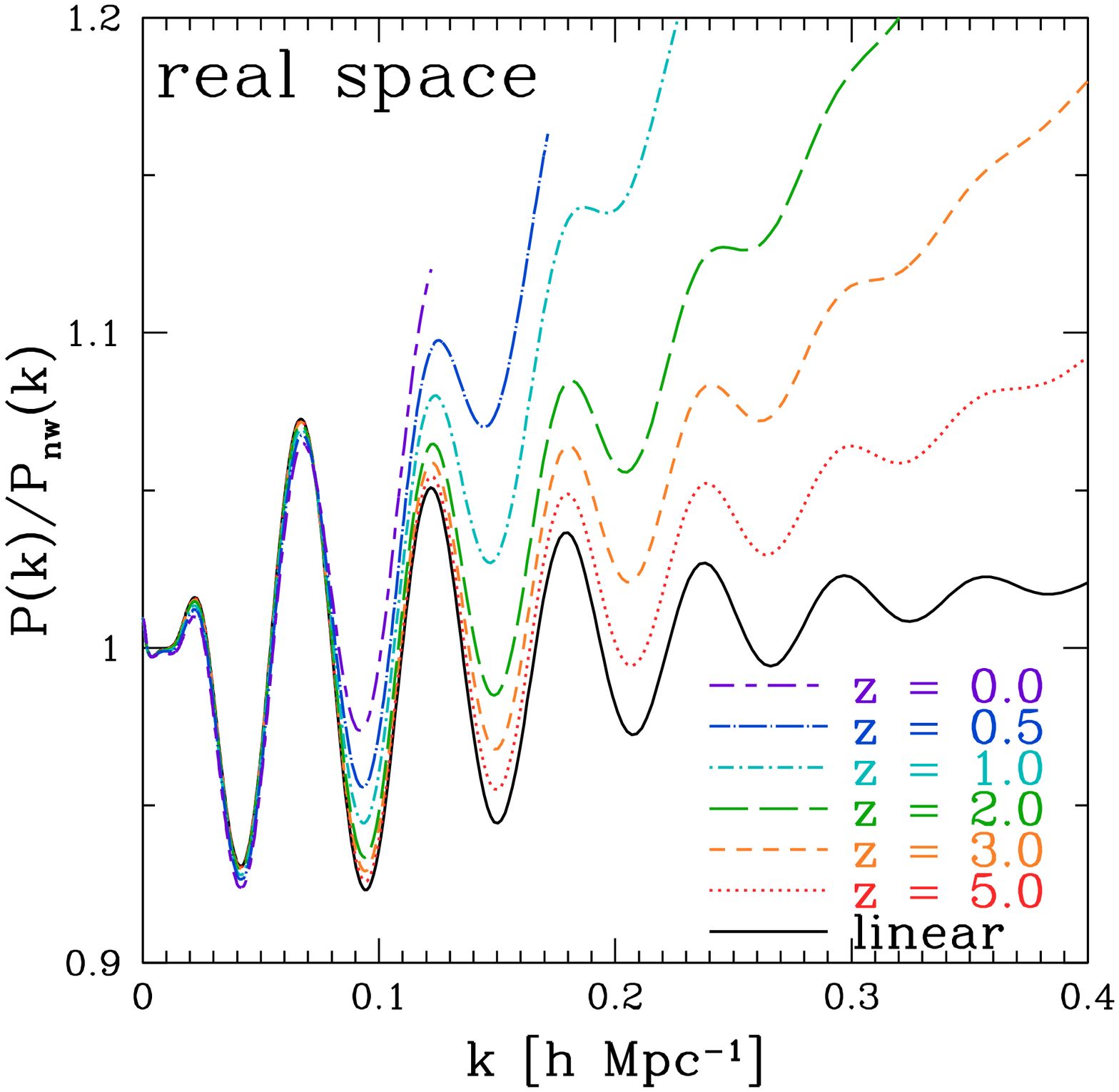}
   \centering \includegraphics[width=0.4\textwidth]{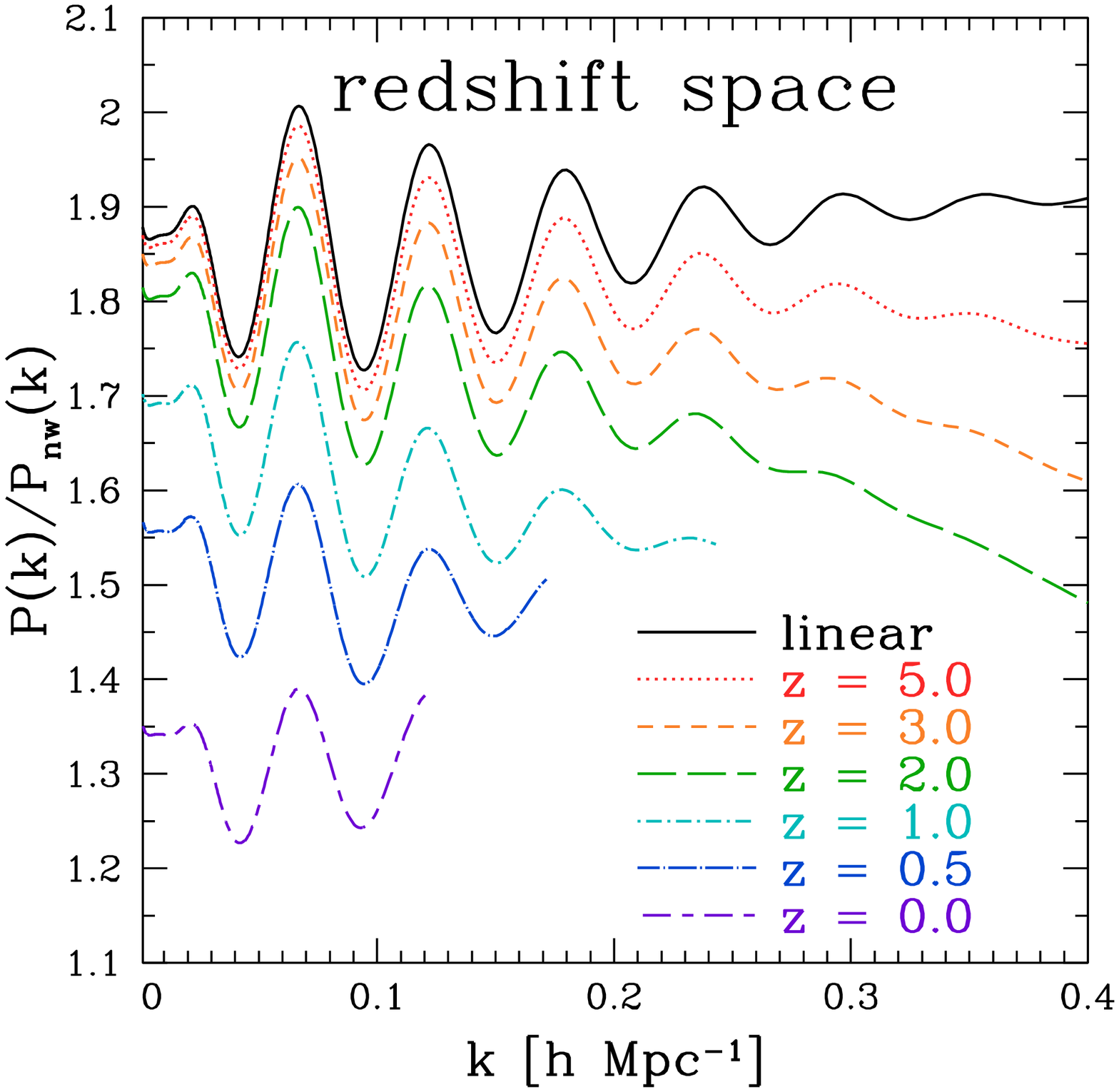}
\caption{The power spectrum divided by no-wiggles approximation, 
$f_{\rm BAO}^{\rm (i)}(k)$, in real (left) and redshift (right) spaces 
(see Eq.[\ref{eq:f1}]). The solid lines represent the results for 
the linear power spectrum. The others indicate the results for one-loop 
power spectrum at redshifts shown in the panels. The results are 
restricted to the range, $k\le k_{1\%}$, where the perturbation theory is 
safely applied (Eq.[\ref{eq:knl}]).
}
\label{fig:pk_over_pknw}
   \centering \includegraphics[width=0.4\textwidth]{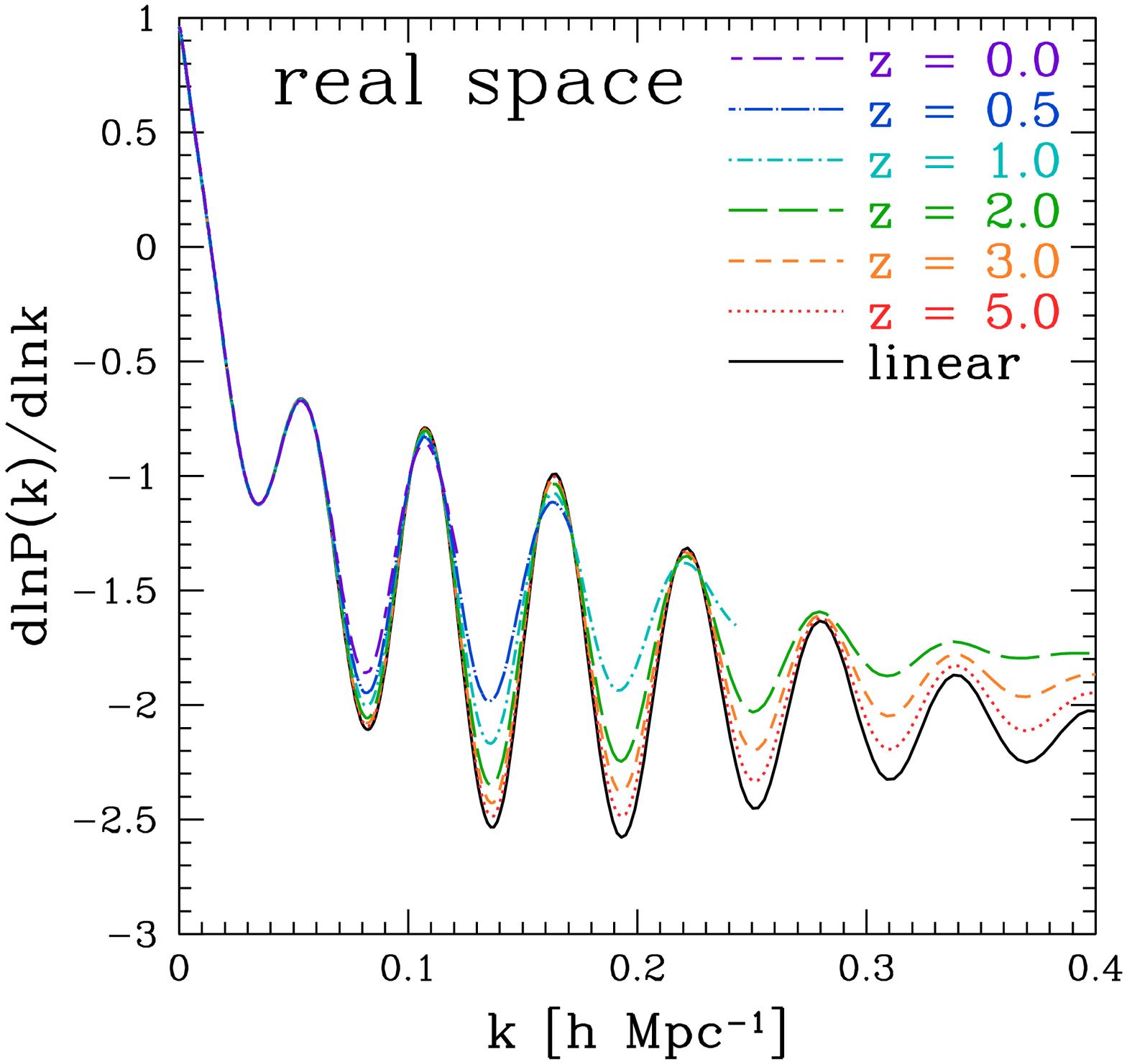}
   \centering \includegraphics[width=0.4\textwidth]{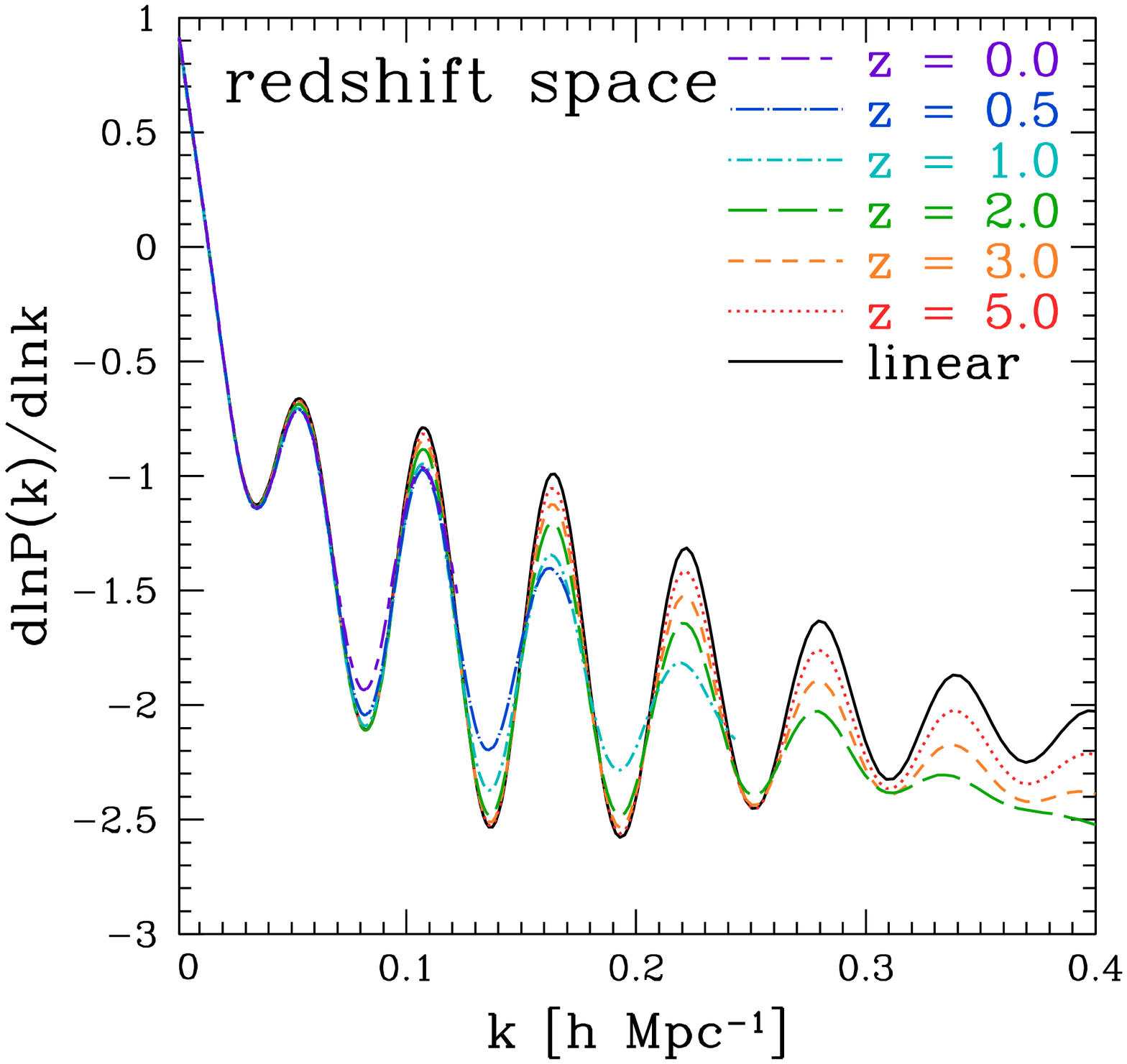}
\caption{Same as figure \ref{fig:pk_over_pknw}, but for $f_{\rm BAO}^{\rm
(ii)}(k)$ 
in equation (\ref{eq:f2}).}
\label{fig:dlnPdlnk}
   \centering \includegraphics[width=0.4\textwidth]{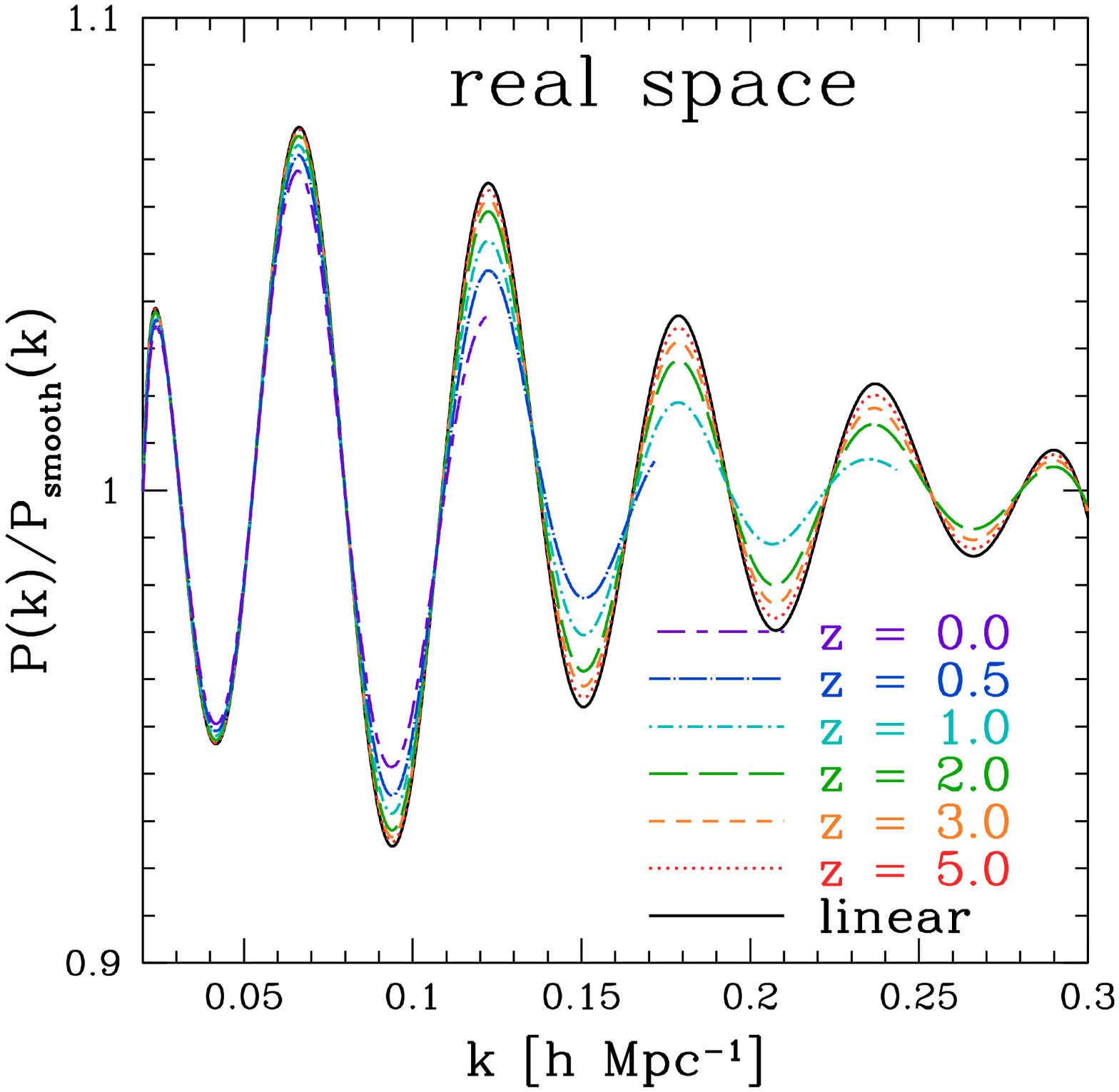}
   \centering \includegraphics[width=0.4\textwidth]{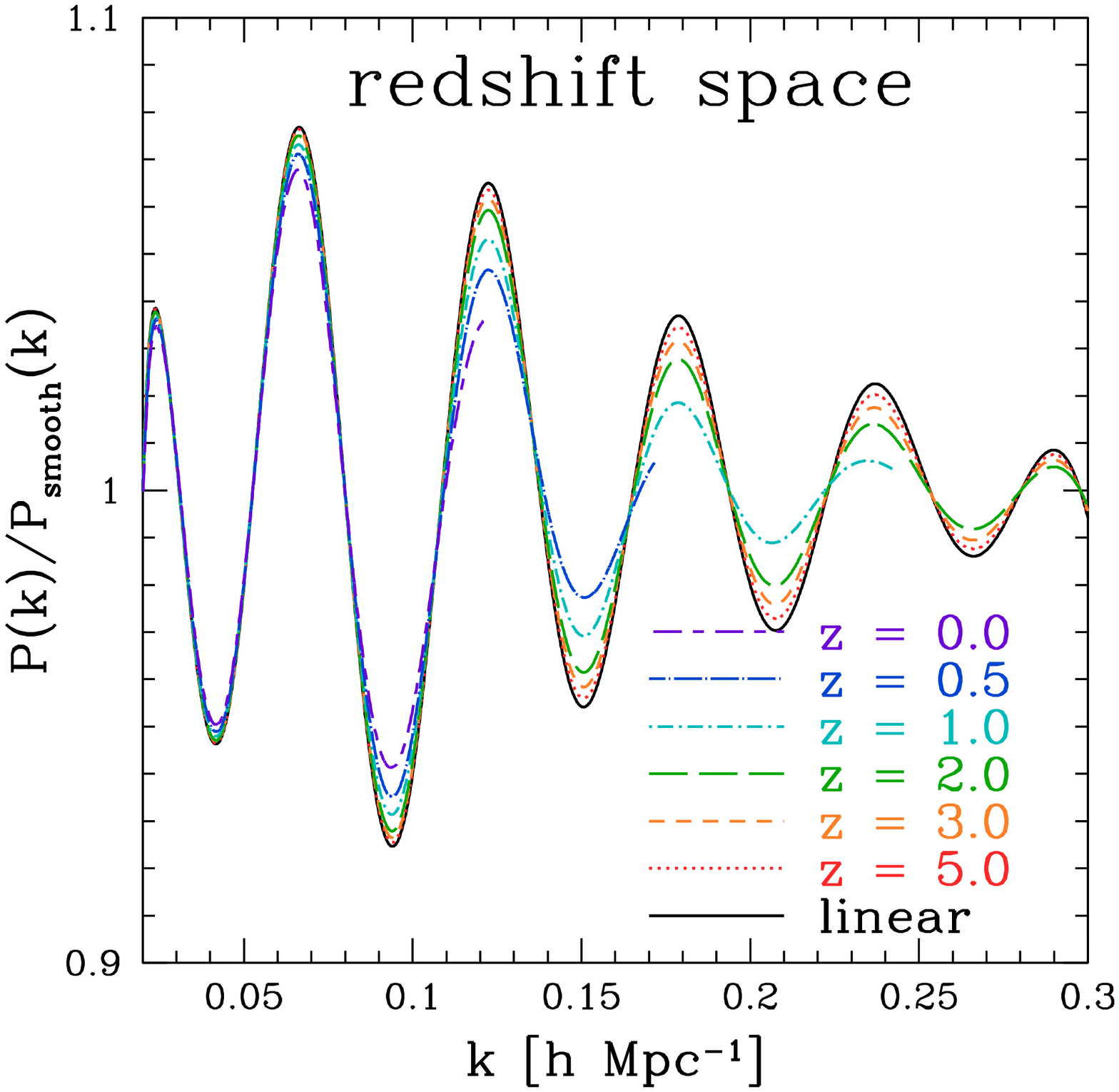}
\caption{Same as figure \ref{fig:pk_over_pknw}, but for 
$f_{\rm BAO}^{\rm (iii)}(k)$ in equation (\ref{eq:f3})}
\label{fig:Percival}
\end{figure}

Now we are in a position to discuss the systematic effects on the BAOs
by evaluating the functions $f_{\rm BAO}(k)$.  Consider first the
general trends of the systematic influences.  In figures
\ref{fig:pk_over_pknw}, \ref{fig:dlnPdlnk} and \ref{fig:Percival},
redshift dependence of the functions $f_{\rm BAO}$ is plotted against
the wave number for three different methods. In each panel, solid lines
represent the results from the linear theory prediction.  Note here that
the plotted curves from the perturbation results are all restricted to
the range, $k\leq k_{\rm 1\%}$, where the perturbation theory is safely
applied (see Eq.[\ref{eq:knl}]).

Basically, depending on the characterization methods, 
the non-linear clustering and the redshift-space distortion 
affect the characteristic scale of BAOs in very different manners. 
In figure \ref{fig:pk_over_pknw}, 
deviations from the linear theory prediction become more significant in 
both real and redshift spaces, as the redshift or scale decreases. 
While the growth of the amplitude $f_{\rm BAO}^{\rm(i)}$ 
is clearly seen in real space, the suppression of the amplitude is 
observed in redshift space, together with the overall 
shifts of acoustic oscillation. The latter effects are 
simply the outcome of the redshift-space distortion. 
By contrast, systematic effects on 
the amplitudes  $f_{\rm BAO}^{\rm(ii)}$ and $f_{\rm BAO}^{\rm(iii)}$, shown 
in figures \ref{fig:dlnPdlnk} and \ref{fig:Percival}, seem rather mild. 
A closer look at the BAO signal reveals that 
the oscillatory features tend to be erased as the redshift 
and the wavenumber increase. This trend 
has also been seen in $N$-body simulations \citep{Seo2005,Angulo2007} 
and modeled by convolving the Gaussian filter with the linear power spectrum 
$P^{\rm L}(k)$ (e.g., \cite{Eisenstein2005}). 
Note that the Gaussian behavior in the disappearance of the 
oscillatory pattern is indeed suggested from the non-perturbative 
prediction based on the renormalized perturbation theory (Crocce \& Scoccimarro
2006b). 
In the present analysis using the perturbation theory, 
the disappearance of the acoustic oscillation mainly comes from 
the one-loop terms $P_{ij}^{(22)}$, showing the monotonic behaviors  
as function of redshift and wave number.

Turn next to focus on the systematic influences on the location of 
the peaks and the troughs. For this purpose, 
we define the fractional shift, $\Delta k/k$, given by 
\begin{eqnarray}
\frac{\Delta k}{k}\equiv 
\left.\frac{k^{\rm PT}-k^{\rm L}}{k^{\rm L}}\right|_{\rm peak\;or\;trough},
\label{eq:frac_shift}
\end{eqnarray}
and quantify the degree of the positional shifts in each method. 
Here, the quantities $k^{\rm PT}$ and $k^{\rm L}$ represent the wave 
number of the peak (or trough) location calculated from the 
perturbation theory and the linear theory, respectively.

\begin{figure}[!ht]
   \centering \includegraphics[width=0.4\textwidth]{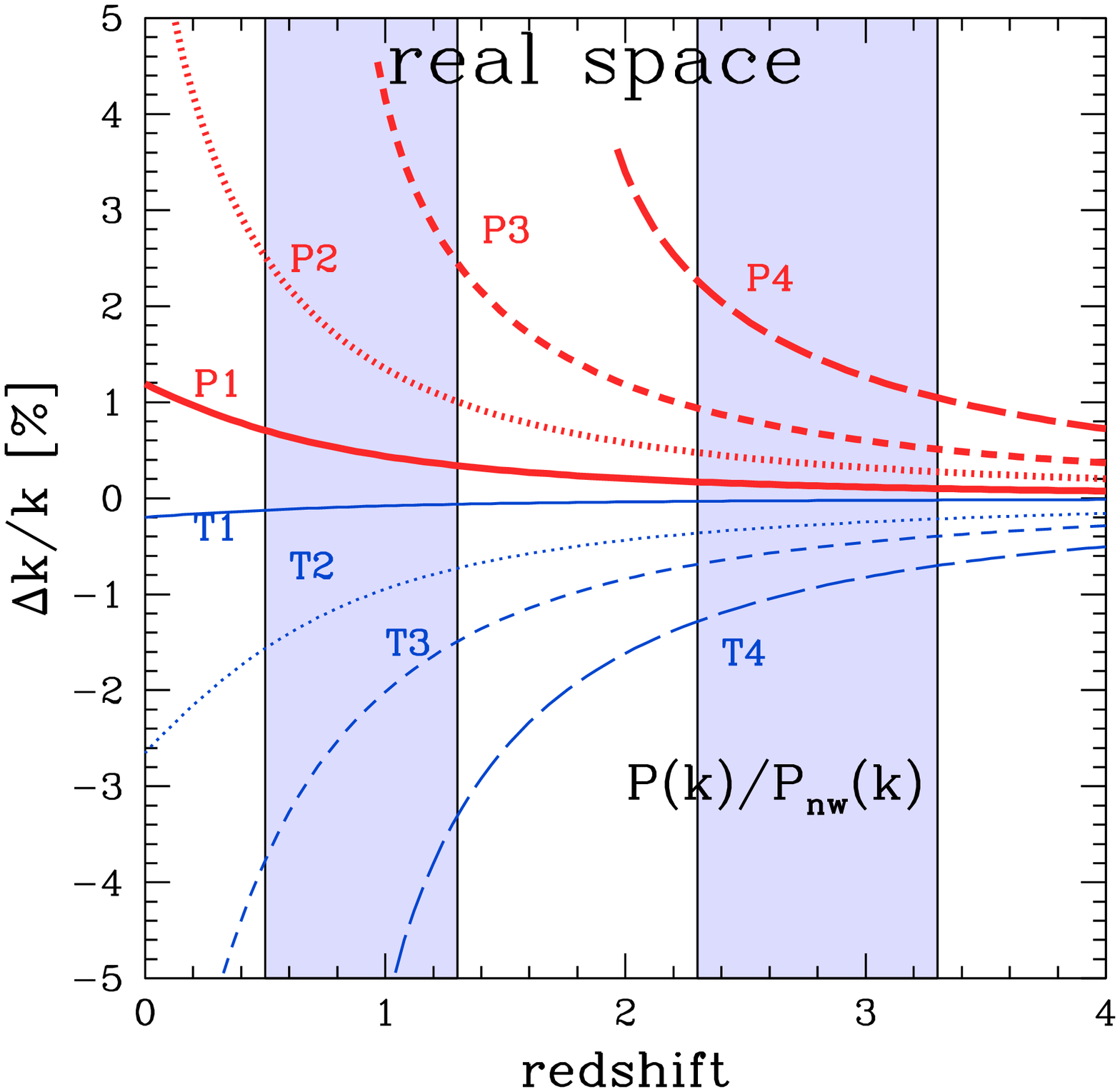}
   \centering \includegraphics[width=0.4\textwidth]{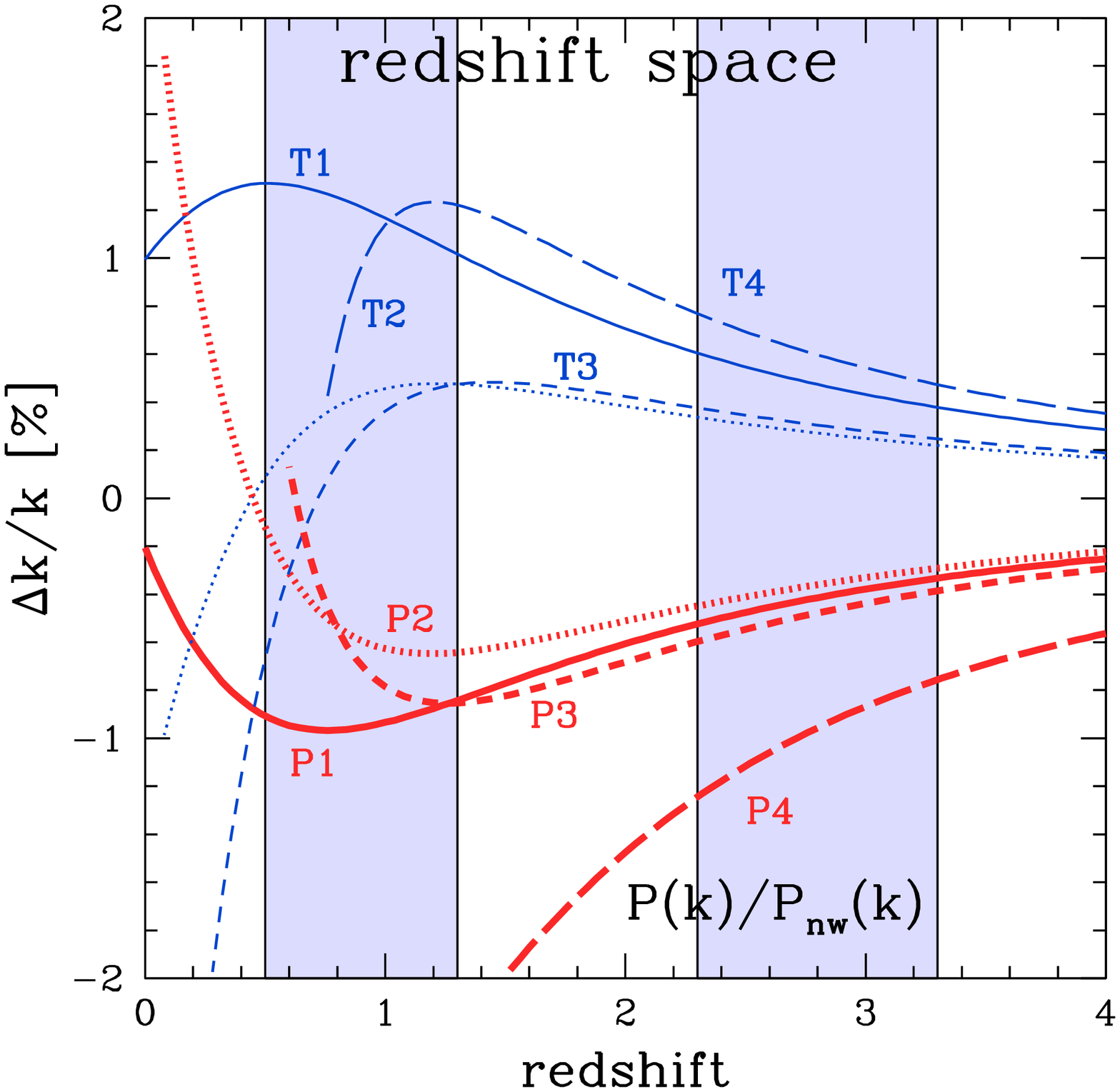}
\caption{The fractional shifts of peaks (from P1 to P4) and troughs 
(from T1 to T4) of $f_{\rm BAO}^{\rm (i)}(k)$. The left (right) 
panel shows the results in real (redshift) space. The two shaded 
regions around $z\sim1$ and $z\sim3$ are the observational windows 
of the planned galaxy redshift survey, WFMOS.}
\label{fig:peak_trough_nw}
   \centering \includegraphics[width=0.4\textwidth]{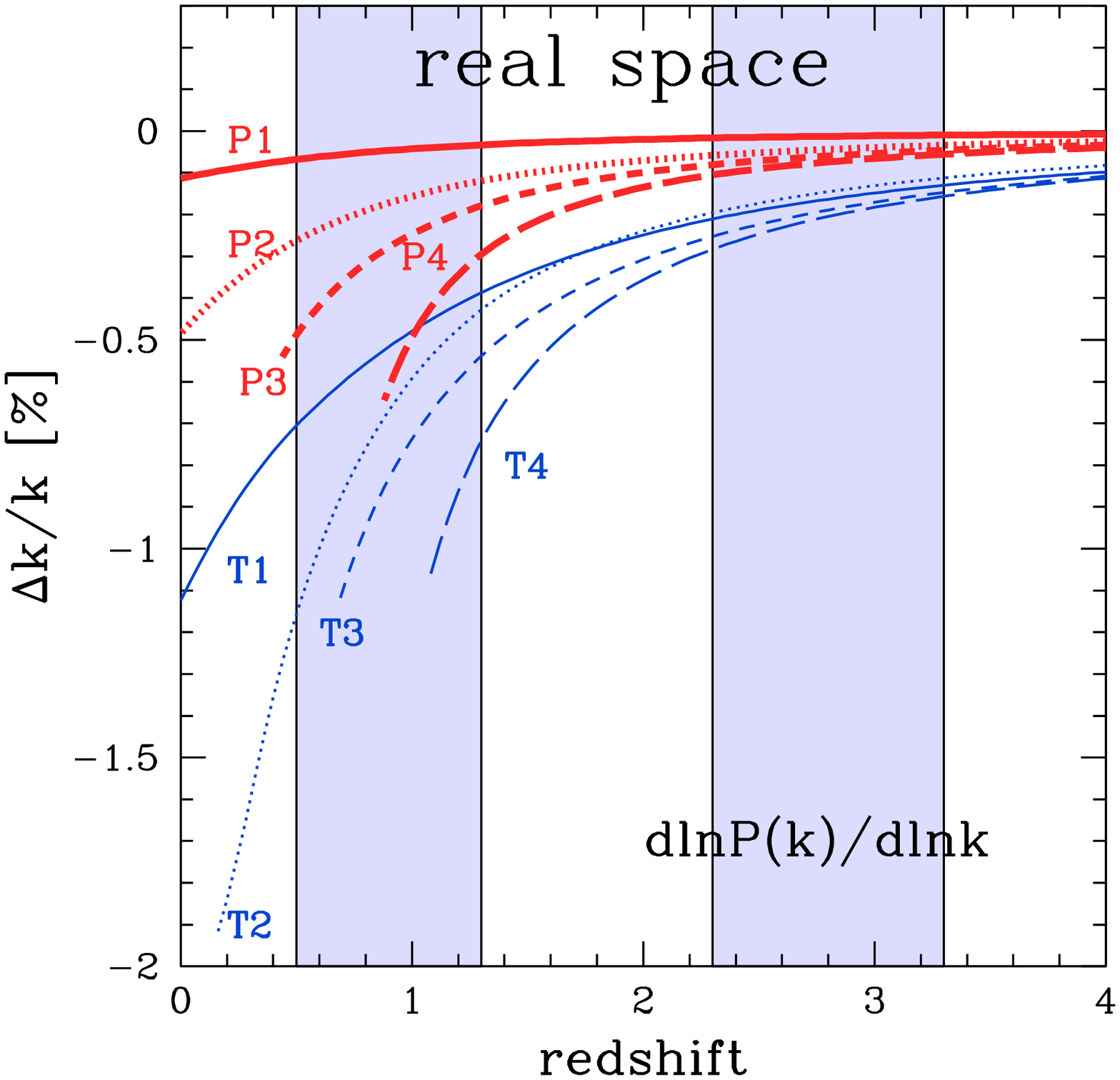}
   \centering \includegraphics[width=0.4\textwidth]{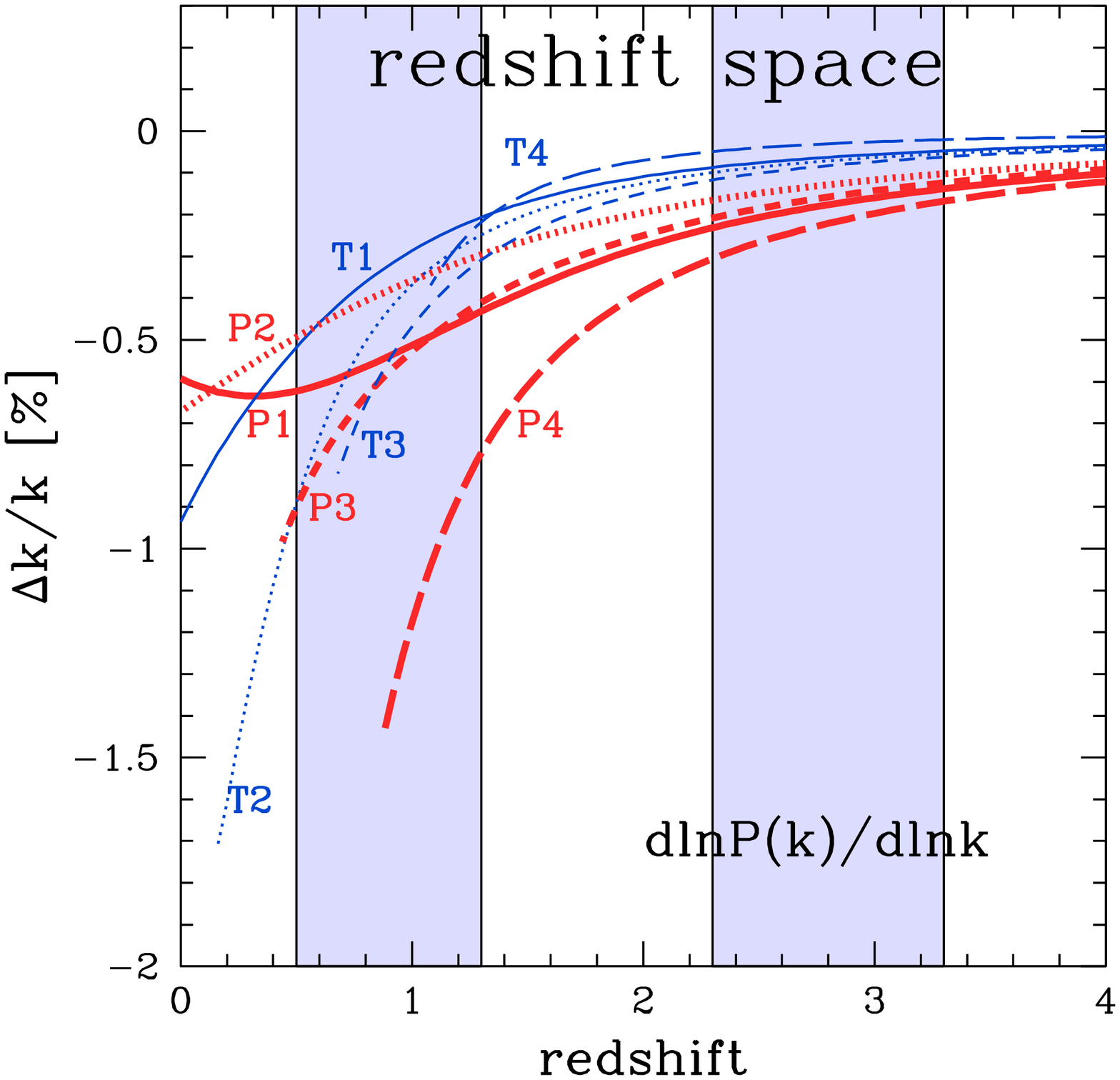}
\caption{Same as figure \ref{fig:peak_trough_nw}, but for 
$f_{\rm BAO}^{\rm (ii)}(k)$.}
\label{fig:peak_trough_dlnPdlnk}
   \centering \includegraphics[width=0.4\textwidth]{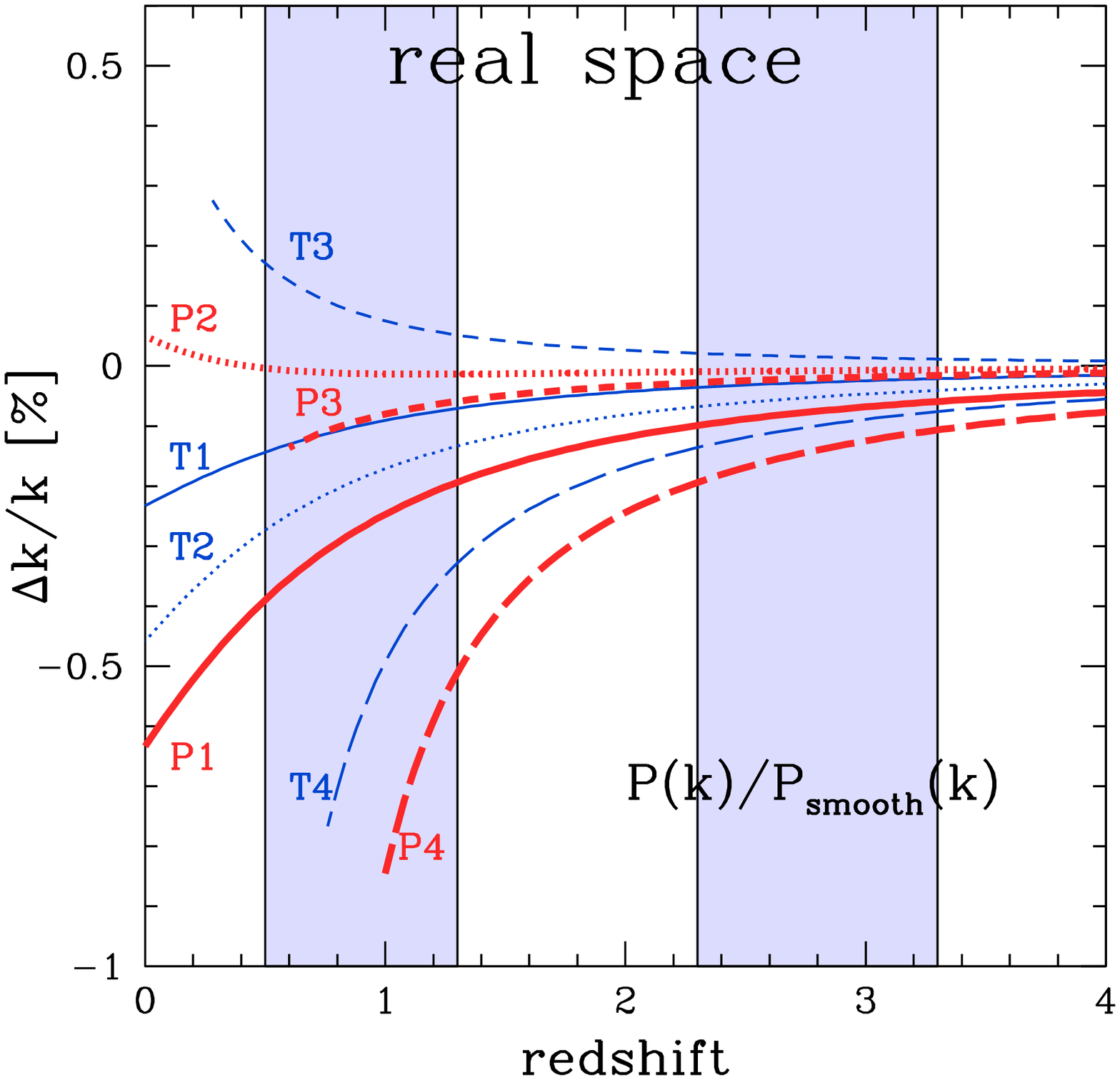}
   \centering \includegraphics[width=0.4\textwidth]{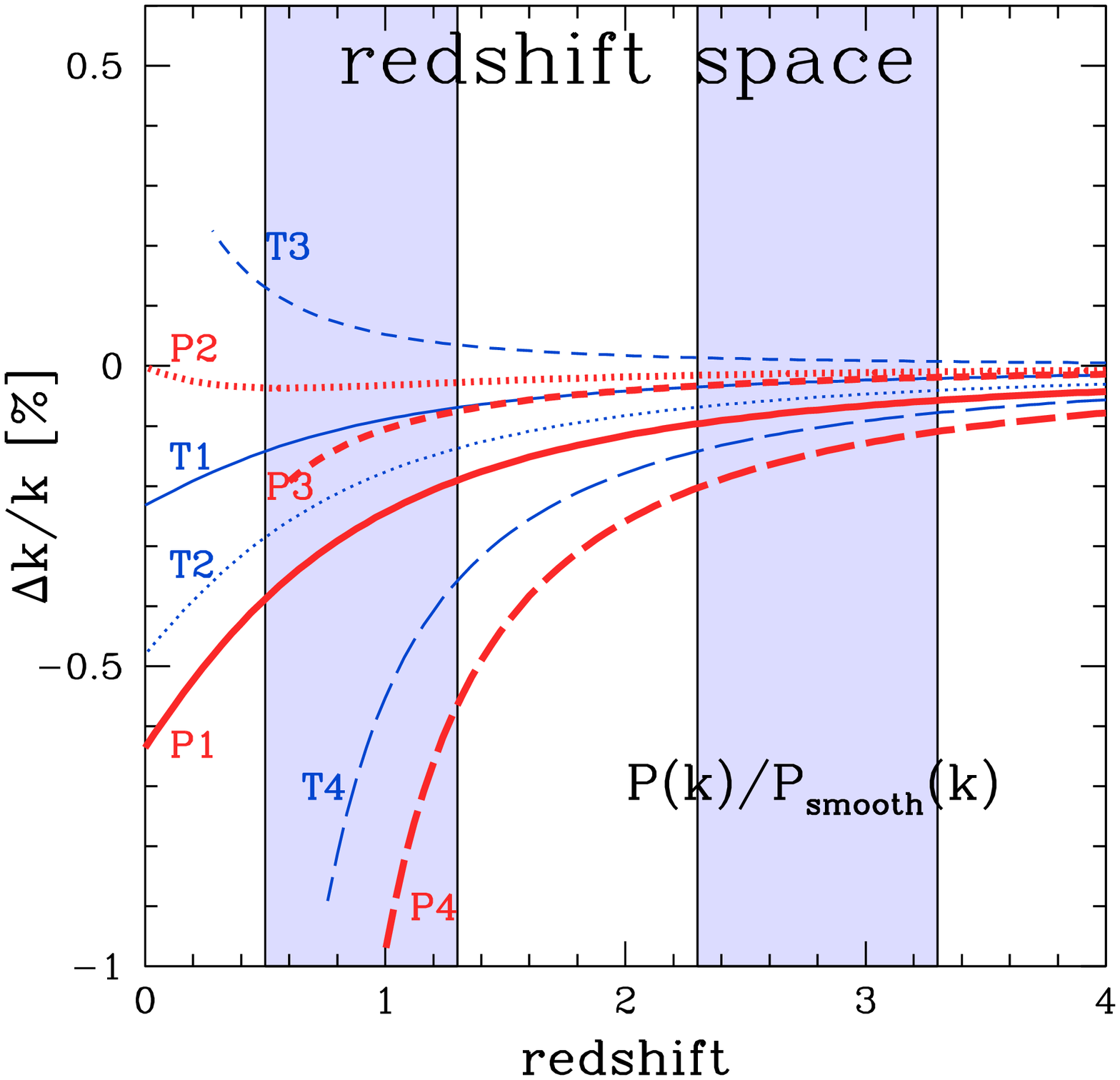}
\caption{Same as figure \ref{fig:peak_trough_nw}, but for 
$f_{\rm BAO}^{\rm (iii)}(k)$.}
\label{fig:peak_trough_Percival}
\end{figure}

Figures \ref{fig:peak_trough_nw}, \ref{fig:peak_trough_dlnPdlnk} and  
\ref{fig:peak_trough_Percival} show the fractional shifts of the peak 
and the trough positions measured from the three different methods,   
plotted as functions of redshift. 
In each panel, thick and thin lines respectively indicate the fractional 
shifts for the peaks (from P1 to P4) and the troughs (from T1 to T4). 
As a reference, we also plot the observational windows of the planned 
galaxy redshift survey, WFMOS, depicted as shaded regions. 
As anticipated from the general trends, the systematic changes in the 
peak and the trough positions become significant on small scales and at
lower redshift. Also, quite naturally, the magnitude of the 
shift depends on the characterization algorithm. At the 
redshift around $z=1$, the fractional shift of the peaks and the troughs 
reaches or exceeds $3\sim4$\% in cases using $f_{\rm BAO}^{\rm(i)}$, 
while in the method $f_{\rm BAO}^{\rm(ii)}$, the systematic effects are 
reduced to $1\sim2$\% level. When translating these results 
to the measurement error of the sound horizon scales, 
the uncertainty in determining the equation of state parameter $w_{\rm DE}$ 
will amount to $12\sim16$\% for the method using 
$f_{\rm BAO}^{\rm(i)}$, and to $4\sim8$\% for the method with 
$f_{\rm BAO}^{\rm(ii)}$ (see Fig.~\ref{fig:dwdr}). On the other hand, 
for the method with 
$f_{\rm BAO}^{\rm(iii)}$, a rather small value of the fractional shifts 
(less than $0.5$\%) is obtained. This is very good news for the 
precision measurement of sound horizon scales. The basic reason  
for the negligible shifts is originated from the definition (\ref{eq:f3}) 
itself that the non-linear growth and the FOG effect are incorporated 
into both the numerator and the denominator, by which major sources for the 
peak and the trough shifts are effectively eliminated. 
As a consequence, the redshift dependence of the fractional shifts 
seen in the real space is almost identical to the one in the 
redshift space. By contrast, the characterization methods with 
$f_{\rm BAO}^{\rm(i)}$ or $f_{\rm BAO}^{\rm(ii)}$ show 
different redshift dependence between real and redshift spaces. 
Interestingly, the behaviors of the fractional shifts 
seen in figure \ref{fig:peak_trough_nw} show some symmetries between 
peaks and troughs.  These systematic trends are related to the 
non-linear corrections arising from the gravitational clustering and 
the redshift-space distortion. We will discussed this issue in some 
detail in the next section.


\section{A toy model for peak and trough shifts}
\label{sec:Interpretation}

The previous section reveals that 
both the non-linear clustering and the redshift-space distortion can lead 
to the physical shift of the peak and trough positions. 
However, their influences apparently depend on the characterization 
method for the BAO signals. In this section, we discuss how to 
understand the positional shifts by introducing a simple toy model.

Our primary goal is a qualitative understanding of the behaviors of 
the positional shift. Let us suppose that the 
acoustic signature in the linear theory prediction can be described by 
a simple sinusoidal function as 
\begin{eqnarray}
f_{\rm BAO}(k) = \sin(k r_s+\delta_0).   
\label{eq:model_lin}
\end{eqnarray}
Of course, this is too naive an assumption, but the outcome of the following 
analysis still keeps the essence of our findings. 
From equation (\ref{eq:model_lin}), the position of the peaks and the 
troughs, denoted by $k^{\rm L}$, satisfies the following relation: 
\begin{eqnarray}
k^{\rm L}r_s+\delta_0 = \left(n+\frac{1}{2}\right)\pi,
\quad(n=0,1,\cdots).
\label{eq:k_p}
\end{eqnarray}
Note that the peak (trough) implies that $n$ is even (odd) number. 
We then add the corrections due to the non-linear 
clustering and the redshift-space distortion, and discuss how the 
correction induces the positional shifts.  
Based on figures \ref{fig:pk_over_pknw}, \ref{fig:dlnPdlnk} and 
\ref{fig:Percival}, the non-linear corrections may be modeled by 
\begin{eqnarray}
f_{\rm BAO}(k) = e^{-(\lambda\,k)^2}\,\sin(k r_s+\delta_0) + A(k).    
\label{eq:model_nl}
\end{eqnarray}
In the above expression, the Gaussian factor 
$e^{-(\lambda k)^2}$ represents 
the smoothing of the acoustic oscillations by the 
non-linear evolution. 
The function $A(k)$ is assumed to be a monotonic function, 
which mimics the residuals that cannot be 
absorbed by taking logarithmic derivative and/or division 
by the smooth spectrum. Roughly speaking,  
monotonically increasing behavior of the function $A(k)$ arises 
from the non-linear growth of gravitational clustering, 
while the monotonically decreasing behavior appears due to the 
FOG effect in redshift space. Note that the parameter $\lambda$ and the 
function $A(k)$ also depend on the redshift.

In the present analysis using perturbation theory, 
the corrections appearing in equation (\ref{eq:model_nl}) should 
be small and perturbative treatment is always valid. 
We express the peak and the trough position by 
$k_*=k^{\rm L}+\Delta k$ and the shift $\Delta k$ is treated as 
small quantity compared to the sound horizon scales, $1/r_s$. 
From the definitions of peaks and troughs, we have  
(see Eq.[\ref{eq:def_peak_trough}]),   
\begin{eqnarray}
&&\left.\frac{d\,f_{\rm BAO}(k)}{dk}\right|_{\rm peak\;or\;trough}\nonumber\\
&&=\left\{-2\lambda^2\,k_*\,\sin(k_*r_s+\delta_0)+
r_s\,\cos(k_*r_s+\delta_0)\right\}
\,e^{-(\lambda\,k_*)^2}+A'(k_*)=0.   
\end{eqnarray}
With a help of the relation (\ref{eq:k_p}),  
the above equation is reduced to the 
expression for the shift $\Delta k$, the result of which     
is summarized as the fractional shift (see Eq.[\ref{eq:frac_shift}]):  
\begin{eqnarray}
\frac{\Delta k}{k}\equiv\frac{k_*-k^{\rm L}}{k^{\rm L}}
\simeq -2\left(\frac{\lambda}{r_s}\right)^2 \,\pm\,
\frac{A'(k^{\rm L})}{k^{\rm L}r_s},\quad
\left\{
\begin{array}{lcl}
+ &:& \mbox{peak}
\\
\\
- &:& \mbox{trough}
\end{array}
\right.,  
\label{eq:model_shift}
\end{eqnarray}
where we have used the fact that 
$(\lambda k^{\rm L})^2\ll k^{\rm L}r_s$ and $k^{\rm L}A'\ll k^{\rm L}r_s$.

\begin{figure}[!ht]
\centering \includegraphics[width=0.4\textwidth]{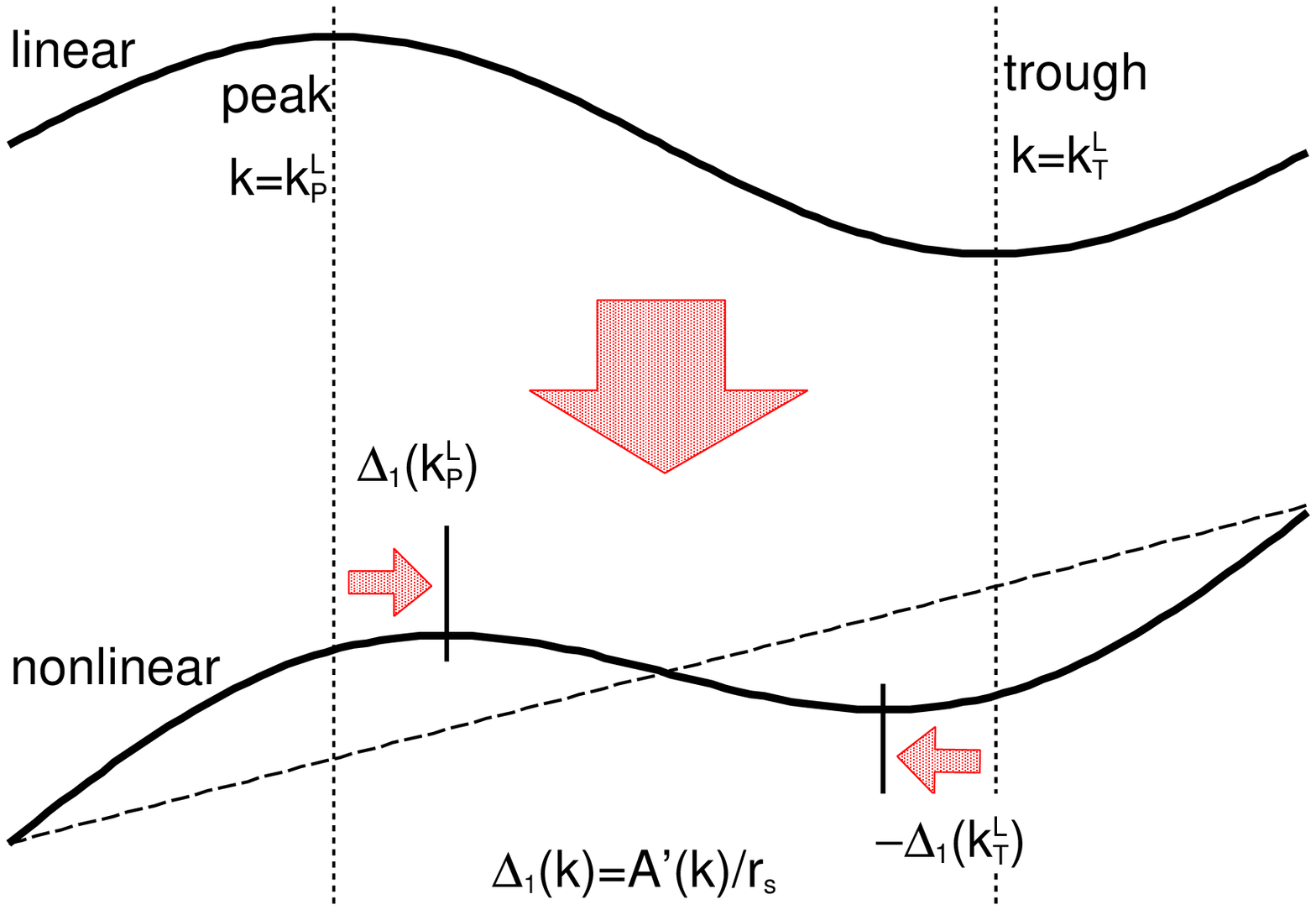}
\hspace*{0.5cm}
\centering \includegraphics[width=0.4\textwidth]{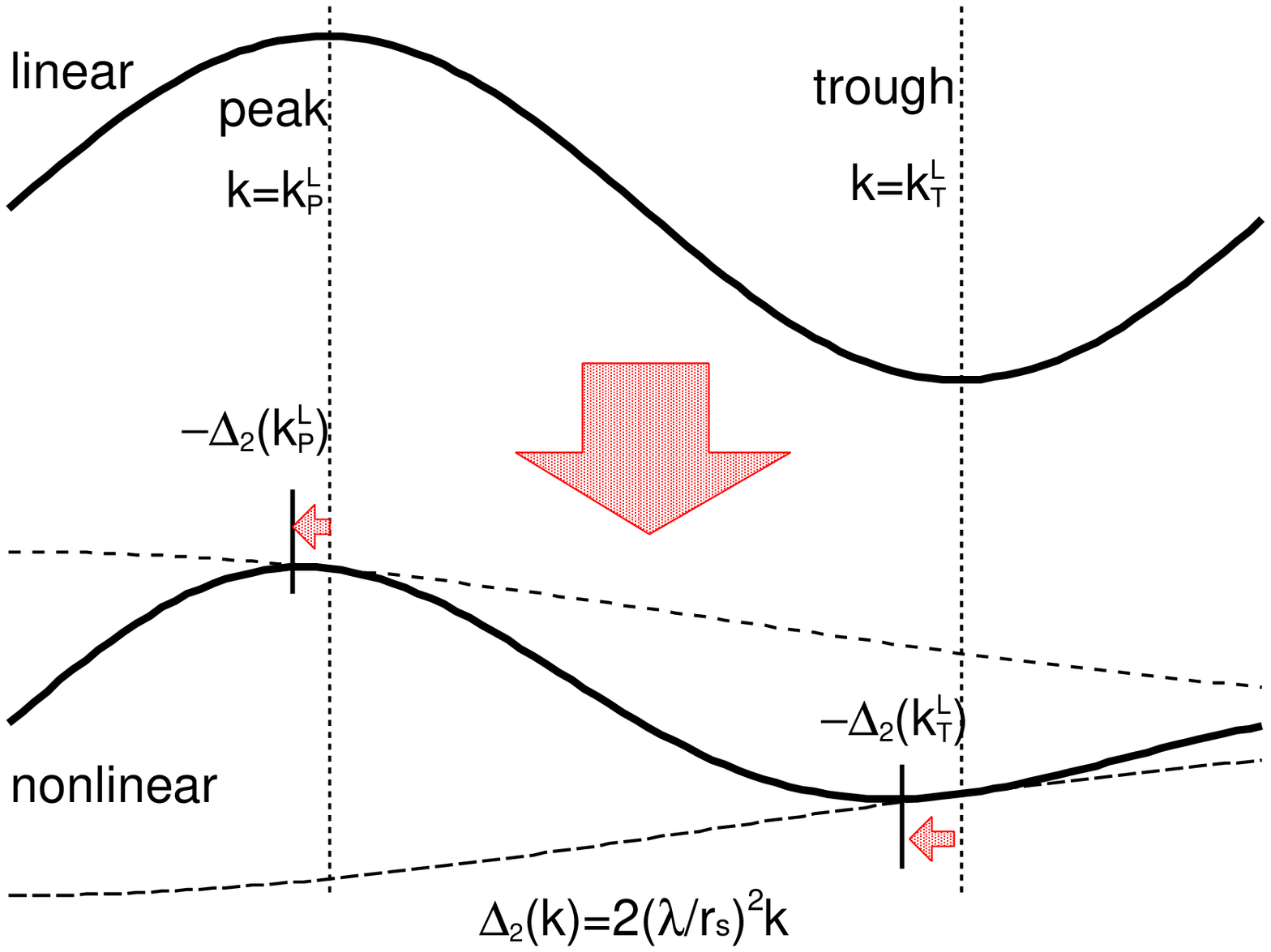}
\caption{Schematic figures of the two reasons for the shifts. 
The upper two curves show $f_{\rm BAO}(k)$ in case of $A'(k)=\lambda=0$, 
while the lowers represent those for $A'(k)>0$ (left) and $\lambda>0$ 
(right). The vertical thin dotted lines represent the peak and the trough 
positions for upper curves, while the short vertical solid lines mark those 
for lower curves. The shifts are given approximately by equation 
(\ref{eq:model_shift}).}
\label{fig:reason}
\end{figure}

From equation (\ref{eq:model_shift}), 
systematic changes in the positional shifts may be interpreted 
as a result of the two competing effects. 
The first term, arising from the Gaussian smoothing factor, always 
makes the position of peaks and troughs move toward smaller $k$. On the other 
hand, the second term, coming from the monotonic behavior of the 
residual corrections, affects the positional shifts symmetrically: 
while the peak moves to the high-$k$ direction, the trough is shifted to 
a small $k$. The magnitude of these trends will be
illuminated more as non-linear corrections become important. 
In figure \ref{fig:reason},  
the role of the two competitive effects are illustrated schematically.

Equation (\ref{eq:model_shift}) qualitatively explains the behaviors seen 
in the previous section. In the case of the function $f_{\rm BAO}^{\rm(i)}$, 
the growth or the suppression of amplitudes was significant and the 
smoothing effect of acoustic signature was sub-dominant 
(Fig.~\ref{fig:pk_over_pknw}). As a result, 
the peaks and the troughs mutually move in an opposite direction, 
consistent with the toy model (\ref{eq:model_shift}). A closer look at the 
late-time evolution in redshift space shows somewhat curious behavior 
that the time evolution of positional shift eventually changes 
its direction from smaller $k$ to larger $k$ for peaks, and 
from larger $k$ to smaller $k$ for troughs (Fig.~\ref{fig:peak_trough_nw}). 
Perhaps, this might result from the imbalance of the two competitive 
effects:  non-linear growth of gravitational clustering and suppression 
by the FOG effect. Hence, if we allow the sign of $A'(k)$ to 
change, this is also explained by the toy model (\ref{eq:model_shift}).  
On the other hand, for the method using $f_{\rm BAO}^{\rm(ii)}$ and 
$f_{\rm BAO}^{\rm(iii)}$, disappearance of acoustic oscillation is the 
major effect (see Figs.~\ref{fig:dlnPdlnk} and \ref{fig:Percival}). 
Although the acoustic signature seen in the function 
$f_{\rm BAO}^{\rm(ii)}$ is primarily declined, centered around 
$d\ln P/d\ln k\simeq-1.7$, this does not essentially affect the 
positional shift. As a consequence, the position of the peaks and the troughs 
systematically moves to the low-$k$ direction. 
Again, we emphasize the remarkably small shift found in the 
function $f_{\rm BAO}^{\rm(iii)}$ (Fig.~\ref{fig:peak_trough_Percival}). 
This implies that the 
corrections corresponding to the $A(k)$ term are 
completely eliminated. Hence, with the characterization method 
$f_{\rm BAO}^{\rm(iii)}$, systematic error in the 
measurement of sound horizon scale would be greatly reduced, leading to 
an accurate determination of the equation of state parameter 
$w_{\rm DE}$.

\section{Conclusions}
\label{sec:Conclusion}

We have considered the shifts of the BAO characteristic scales due to
the nonlinear gravitational and redshift-space distortion effects in a
weakly nonlinear regime using one-loop correction from perturbation
theory.  We adopted three different methods to define the BAO
oscillatory features from the entire power spectrum, and compute the
shifts of peak and trough locations relative to the purely linear theory
predictions in real space.

In doing so, we presented an analytic toy model to account for the
physical reasons for the shifts, and showed that one particular method
similar to the earlier proposal by \citet{Percival2007} is fairly free
from the nonlinear and redshift-space distortion effects.  In practice,
the shifts of the first few peak and trough locations defined in the
above procedure are at the ${\mathrel{\raise1.2pt\hbox{$<$}\kern-8.8pt
\lower3.2pt\hbox{{$\sim$}}}}0.5\%$ level, ensuring ${\mathrel{\raise1.2pt
\hbox{$<$}\kern-8.8pt\lower3.2pt\hbox{{$\sim$}}}}2\%$ precision in terms 
of the dark energy parameter $w_{\rm DE}$, {\it even
if one uses the linear theory predictions as a standard ruler}. Of
course the shifts can be accurately computed using our methodology as
long as the one-loop correction is dominant in the regime of interest.

The result is fairly robust against possible additional effects such as
a weakly scale-dependent biasing and a running spectral index because
they preferentially change the smoothed component in the power spectrum
that is almost removed from the above procedure. 

Of course the next task is to establish an accurate model that predicts
the amplitude of the BAO under the nonlinear and redshift-space
distortion effects. We suspect that it is very challenging given the
limitation in both the current theoretical framework and numerical
simulations. In particular, the nonlinear stochastic nature of the galaxy 
biasing seems problematic (e.g., \cite{Taruya2000,Nishimichi2007}). 
It is still very difficult to incorporate this effect into theoretical 
predictions in any realistic and believable manner. 

In light of this, it may be reasonable at this point to use 
the BAO scale information exclusively in constraining dark energy,
ignoring its amplitude. In practice, constraints on 
cosmological parameters from the use of the BAO scale as a standard 
ruler hinges on the feasibility of the simultaneous fitting of the 
multiple BAO peaks and troughs. Furthermore, two-dimensional 
(line-of-sight and plane of the sky directions) features in 
redshift-space may improve the accuracy on the BAO scale. 
We are currently working on a simulation-based study, which is 
necessary to investigate these issues quantitatively.

\bigskip

We thank M. Takada, A. Nishizawa and E. Reese for useful comments 
related to the topic in the present paper. This work is supported 
in part by Japan Society for Promotion of Science (JSPS) Core-to-Core 
Program ``International Research Network for Dark Energy''. T.N, A.S and 
K.Y acknowledge the support from the JSPS Research Fellows. 
A.T and K.Y are supported by a a Grant-in-Aid for Scientific Research 
from the JSPS (Nos.~18740132,~18540277,~18654047). 

\appendix
\def\thesection{\Alph{section}}
\def\theequation{\thesection\arabic{equation}}

\section{The linear theory prediction of the characteristic scales}
\label{app:phase}

The BAO characteristic scale imprinted in the matter power spectrum is
basically the sound horizon scale at recombination, $r_s(z_{\rm rec})$
(eq.[\ref{eq:soundhorizon}]). Depending on the specific definitions of
peaks and troughs in $k$-space that we adopted here, however, their
corresponding scales are slightly different from the value of equation
(\ref{eq:soundhorizon}), which has a non-negligible effect in estimating
the cosmological parameters. The purpose of this Appendix is to clarify
the difference in linear theory predictions.

The observed BAO scale in real space is expected to differ slightly from
$r_s(z_{\rm rec})$ due to the residual baryon-photon interaction after
recombination ($z_{\rm rec}\approx 1089$).  \citet{Eisenstein1998}
pointed out that a more accurate value is given simply by replacing the
$z_{\rm rec}$ with the drag epoch $z_d$:
\begin{eqnarray}
\label{eq:rs-zd}
r_s(z_{d}) = \frac{2}{3k_{\rm eq}}\sqrt{\frac{6}{R_{\rm eq}}}\ln
\frac{\sqrt{1+R_d}+\sqrt{R_d+R_{\rm eq}}}{1+\sqrt{R_{\rm eq}}},
\end{eqnarray}
where $R_d=R(z_d)$ is the ratio of the baryon to photon momentum
densities at $z_d=1019$. Equation (\ref{eq:rs-zd}) implies that
$r_s(z_d)=155$Mpc, which is about 5\% larger than $r_s(z_{\rm
rec})=148$Mpc.

The characteristic scale in k-space is even more subtle. So let us first
model the oscillating part of baryon transfer function as 
\begin{eqnarray}
T_b(k) \propto \sin\phi.
\end{eqnarray}
If we adopt equation (\ref{eq:soundhorizon}), the phase $\phi$ is written
as
\begin{eqnarray}
\phi=kr_s(z_{\rm rec}).
\label{eq:phase_rec}
\end{eqnarray}
\citet{Eisenstein1998} took into account of baryon density perturbation
at the drag epoch itself, which changes the phase at large scales where
the velocity overshoot is not the dominant effect. As a result, they 
found that the phase $\phi$ is approximated as
\begin{eqnarray}
\phi = k\tilde{r}_s(k),
\label{eq:phase_EH}
\end{eqnarray}
where
\begin{eqnarray}
\tilde{r}_s(k) &=& \frac{r_s(z_d)}{[1+(\beta_{\rm node}/kr_s(z_d))^3]^{1/3}},\\
\beta_{\rm node} &=& 8.41(\Omega_mh^2)^{0.435}.
\label{eq:EHsoundhorizon}
\end{eqnarray}
In these expressions, the characteristic scales in $k-$space are defined
through
\begin{eqnarray}
\phi = \frac{\pi}{2}m,
\end{eqnarray}
where $m=5,9,13,17,...$ for peaks $m=3,7,11,14,...$ for troughs from our
methods (i) and (iii). In contrast, our method (ii) implies that
$m=4,8,12,16,...$ for peaks $m=6,10,14,18,...$ for troughs.

\begin{figure}[!ht]
\centering \includegraphics[width=0.5\textwidth]{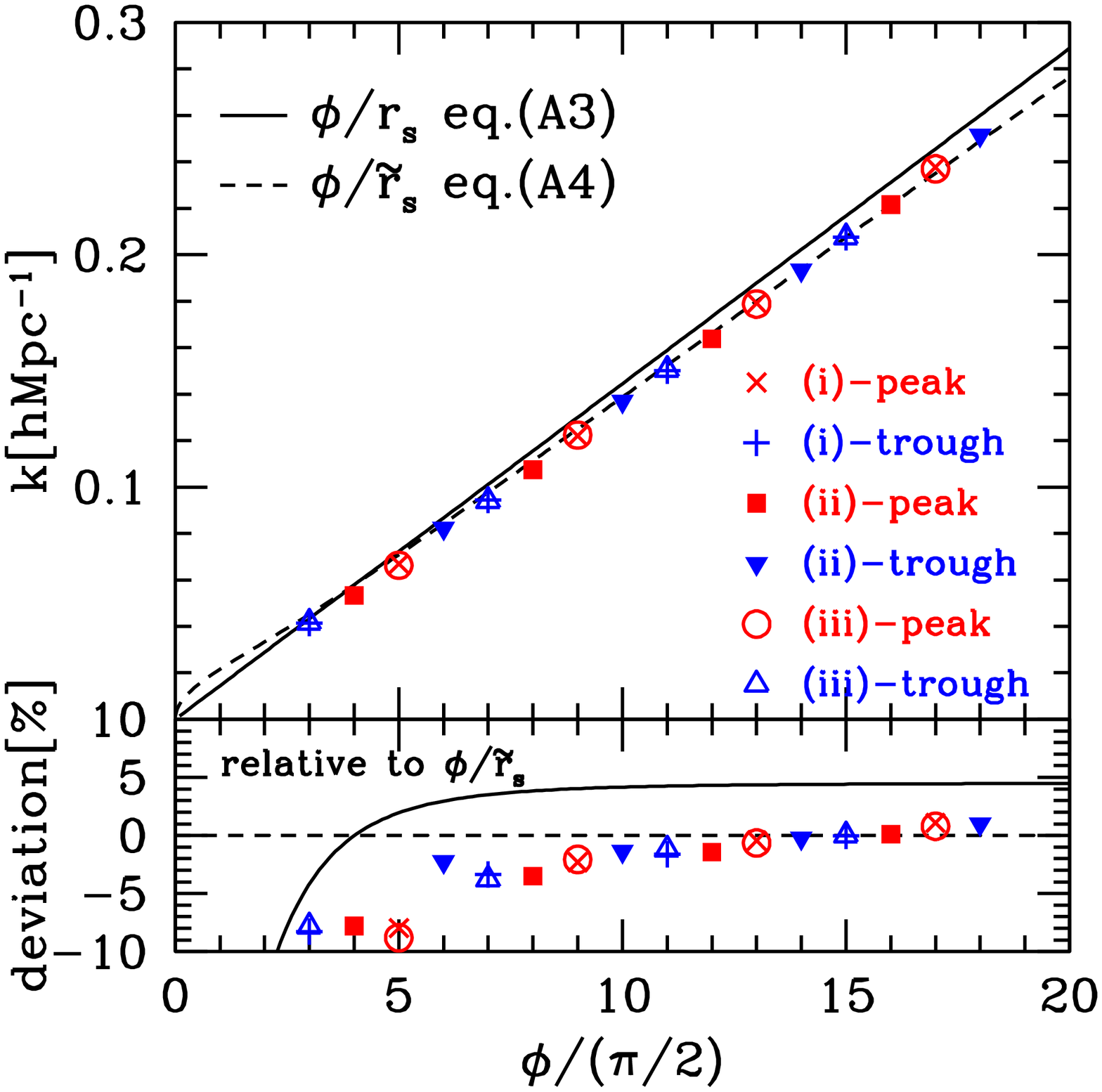}
\caption{The positions of peaks and troughs predicted from linear theory.}
\label{fig:phase}
\end{figure}

Figure \ref{fig:phase} compares those theoretical predictions with our
peak and trough positions in Table \ref{tab:linear_position} (computed
using {\tt CAMB}).  The upper panel shows the relation between the phase
$\phi$ and the wavenumber $k$; the solid and dashed lines are calculated
using equation (\ref{eq:phase_rec}) and (\ref{eq:phase_EH}),
respectively. The symbols denote our results for peaks and troughs with
the three methods listed in Table \ref{tab:linear_position}.  We also
plot the fractional deviation with respect to equation
(\ref{eq:phase_EH}) in the lower panel.

The solid and dashed lines suggest that equations (\ref{eq:phase_rec})
and (\ref{eq:phase_EH}) are indeed different by 5\% as expected.  This
is crucial since the difference propagates to $\sim$20\% in $w_{\rm
DE}$. The dashed line fits our {\tt CAMB} results (symbols) very well
for $\phi>5\pi$, which ensures the validity of the formula of
\citet{Eisenstein1998}. Nevertheless our results for the first few peaks
and troughs are systematically different from their fitting formula (the
lower panel of Figure \ref{fig:phase}). This would simply reflect the
difference of our {\it definitions} of the peaks and troughs with
respect to theirs, which one should keep in mind when performing an
actual statistical analysis of real datasets.

\section{Kernels for perturbative solution and one-loop power 
spectrum}
\label{app:Kernel}
\setcounter{equation}{0}
Here, we briefly summarize the kernels for perturbative solutions 
presented in equation (\ref{eq:Kernel}) and present the explicit 
expressions for the one-loop power spectrum (see Eqs.[\ref{eq:P13}] and 
[\ref{eq:P22}]).

In the Einstein-de Sitter universe, the kernel 
${\cal F}^{(n)}_a({\bf q}_1,\cdots,{\bf q}_n)$ satisfies the following 
recursion relation \citep{Goroff1986}:
\begin{eqnarray}
&&{\cal F}_a^{(1)}({\bf q}_1) = (1,1),\\
&&{\cal F}_a^{(n)}({\bf q}_1,\cdots,{\bf q}_n) = \sigma_{ab}(n)
\sum_{m=1}^{n-1}\gamma_{bcd}({\bf k},{\bf k}_1,{\bf k}_2)
{\cal F}_c^{(m)}({\bf q}_1,\cdots,{\bf q}_m)\nonumber\\
&&\hspace*{3.5cm}\times{\cal F}_d^{(n-m)}({\bf q}_{n-m+1},\cdots,{\bf q}_n),
\end{eqnarray}
where ${\bf k}\equiv{\bf q}_1+\cdots+{\bf q}_n$, ${\bf k}_1\equiv
{\bf q}_1+\cdots+{\bf q}_m$, ${\bf k}_2={\bf q}_{m+1}+\cdots+
{\bf q}_n$, and 
\begin{eqnarray}
\sigma_{ab}(n) = \frac{1}{(2n+3)(n-1)}\left[
\begin{array}{cc}
2n+1 & 2\\
3 & 2n
\end{array}
\right].
\label{eq:sigmaab}
\end{eqnarray}
From these relations,  one obtains the symmetrized kernels, $F^{(n)}_s$ 
(density part) and $G^{(n)}_s$ (velocity part) (e.g., \cite{Jain1994}):
\begin{eqnarray}
F_s^{(2)}({\bf{q}}_1,{\bf{q}}_2)&=&\frac{5}{7}+\frac{1}{2}
\frac{{\bf q}_1\cdot{\bf q}_2}{q_1q_2}\left(\frac{q_1}{q_2}
+\frac{q_2}{q_1}\right)+\frac{2}{7}\frac{({\bf q}_1
\cdot{\bf q}_2)^2}{q_1^2q_2^2},
 \label{F2}
\end{eqnarray}

\begin{eqnarray}
G_s^{(2)}({\bf{q}}_1,{\bf{q}}_2)&=&\frac{3}{7}+\frac{1}{2}
\frac{{\bf q}_1\cdot{\bf q}_2}{q_1q_2}\left(\frac{q_1}{q_2}
+\frac{q_2}{q_1}\right)+\frac{4}{7}\frac{({\bf q}_1
\cdot{\bf q}_2)^2}{q_1^2q_2^2},
 \label{G2}
\end{eqnarray}

\begin{eqnarray}
F_s^{(3)}({\bf{q}}_1,{\bf{q}}_2,{\bf{q}}_3)&=&\frac{1}{6}
\Biggl\{\frac{7}{9}\frac{{\bf{k}}\cdot{\bf{q}}_3}{q_3^2}
F_s^{(2)}({\bf{q}}_1,{\bf{q}}_2)+\biggl[\frac{7}{9}
\frac{{\bf{k}}\cdot({\bf{q}}_1+{\bf{q}}_2)}{|{\bf{q}}_1
+{\bf{q}}_2|^2}+\frac{2}{9}\frac{k^2{\bf{q}}_3
\cdot({\bf{q}}_1+{\bf{q}}_2)}{|{\bf{q}}_1+{\bf{q}}_2|^2 q_3^2}
\biggr]G_s^{(2)}({\bf{q}}_1,{\bf{q}}_2)\Biggr\}\nonumber\\
&&+{\rm cyclic},
 \label{F3}
\end{eqnarray}

\begin{eqnarray}
G_s^{(3)}({\bf{q}}_1,{\bf{q}}_2,{\bf{q}}_3)&=&\frac{1}{6}
\Biggl\{\frac{1}{3}\frac{{\bf{k}}\cdot{\bf{q}}_3}{q_3^2}
F_s^{(2)}({\bf{q}}_1,{\bf{q}}_2)+\biggl[\frac{1}{3}
\frac{{\bf{k}}\cdot({\bf{q}}_1+{\bf{q}}_2)}{|{\bf{q}}_1
+{\bf{q}}_2|^2}+\frac{2}{3}\frac{k^2{\bf{q}}_3
\cdot({\bf{q}}_1+{\bf{q}}_2)}{|{\bf{q}}_1+{\bf{q}}_2|^2 q_3^2}
\biggr]G_s^{(2)}({\bf{q}}_1,{\bf{q}}_2)\Biggr\}\nonumber\\
&&+{\rm cyclic},
 \label{G3}
\end{eqnarray}

The explicit expressions for one-loop correction terms in equations 
(\ref{eq:P13}) and (\ref{eq:P22}) are (e.g., \cite{Makino1992})
\begin{eqnarray}
P_{\delta\delta}^{(22)}(k) &=& \frac{k^3}{98(2\pi)^2}\int_0^\infty 
drP^{\rm L}(kr)\int_{-1}^1dxP^{\rm L}\left(k\sqrt{1+r^2-2rx}\right)
\frac{(3r+7x-10rx^2)^2}{(1+r^2-2rx)^2},
\label{eq:Pdd22}\\
P_{\delta\delta}^{(13)}(k) &=& \frac{k^3}{252(2\pi)^2}P^{\rm L}(k)
\int_0^\infty drP^{\rm L}(kr)\nonumber\\
&&\times
\left[\frac{12}{r^2}-158+100r^2-42r^4
+\frac{3}{r^3}(r^2-1)^3(7r^2+2)\ln\left|\frac{1+r}{1-r}\right|\right],
\label{eq:Pdd13}
\end{eqnarray}
for the power spectrum of density field, 
\begin{eqnarray}
P_{\theta\theta}^{(22)}(k)&=&\frac{k^3}{98(2\pi)^2}\int_0^\infty 
dr P^{\rm L}(kr)
\int_{-1}^1 dx P^{\rm L}\left(
k\sqrt{1+r^2-2rx}\right)\frac{(-r+7x-6rx^2)^2}{(1+r^2-2rx)^2},
\label{eq:Ptt22}\\
P_{\theta\theta}^{(13)}(k)&=& \frac{k^3}{84(2\pi)^2}P^{\rm L}(k)
\int dr P^{\rm L}(kr)\nonumber\\
&&\times\left[\frac{12}{r^2}-82+4r^2-
6r^4+\frac{3}{r^3}(r^2-1)^3(r^2+2)\ln\left|
\frac{1+r}{1-r}\right|\right],
\label{eq:Ptt13}
\end{eqnarray}
for the cross power spectrum of density and velocity divergence, and 
\begin{eqnarray}
P_{\delta\theta}^{(22)}(k)&=&\frac{k^3}{98(2\pi)^2}\int_0^\infty
 dr P^{\rm L}(kr)
\int_{-1}^1 dx P^{\rm L}\left(k\sqrt{1+r^2-2rx}\right)
\nonumber\\
&&\times\frac{(3r+7x-10rx^2)(-r+7x-6rx^2)}{(1+r^2-2rx)^2},
\label{eq:Pdt22}\\
P_{\delta\theta}^{(13)}(k)&=& \frac{k^3}{252(2\pi)^2}P^{\rm L}(k)
\int_0^\infty dr P^{\rm L}(kr)\nonumber\\
&&\times\left[\frac{24}{r^2}-202+56r^2
-30r^4+\frac{3}{r^3}(r^2-1)^3(5r^2+4)\ln\left|\frac{1+r}{1-r}\right|
\right],
\label{eq:Pdt13}
\end{eqnarray}
for the power spectrum of velocity divergence.

\bigskip

\end{document}